\documentclass{iopart}
\usepackage{iopams}
\usepackage{bbm}

\def \R{{\mathbb R}}
\def \C{{\mathbb C}}
\def \Z{\mathbb Z}
\def \N{\mathbb{N}}

\def \D{{\mathcal D}}

\def \H{\mathcal{H}}

\def \a{\alpha}

\def \d{\delta}

\def \t{\tau}

\def \n{\nu}

\def \del{\partial}

\newtheorem{Lemma}{Lemma}[section]

\newtheorem{Proposition}{Proposition}[section]

\eqnobysec

\begin{document}
\title[Quantum mechanics on a circle]
{Quantum mechanics on a circle: Husimi phase space distributions
and semiclassical coherent state propagators}
\author{B Bahr$^1$ and H J Korsch$^2$}

\address{$^1$ Max-Planck Institute for Gravitational Physics (Albert-Einstein Institute),
Am M\"uhlenberg 1, D-14476 Golm, Germany}
\address{$^2$ Technical University of Kaiserslautern,
Gottlieb-Daimler-Stra\ss e, D-67663 Kaiserslautern, Germany}

\ead{bbahr@aei.mpg.de}

\begin{abstract}
We discuss some basic tools for an analysis of  one-dimensional
quantum systems defined on a cyclic coordinate space.
The basic features of the generalized coherent states,
the complexifier coherent states are reviewed. These states are
then used to define the corresponding (quasi)densities in phase space.
The properties of these generalized Husimi distributions are
discussed, in particular their zeros.
Furthermore, the use of the complexifier coherent states for a
semiclassical analysis is demonstrated by deriving a
semiclassical coherent state propagator in phase space.
\end{abstract}
\submitto{\JPA}

\section{Introduction}

\noindent Quantum systems depending on a cyclic coordinate appear in
numerous studies. Celebrated examples of such systems are the
particle on a circle, the plane pendulum, the rotor, but also other
systems such as space periodic crystals in a tight-binding
approximation, where the cyclic variable is the quasimomentum
\cite{03TBalg}. Despite of much effort, the theory of quantization
of such systems is still far from complete (see, e.g.,
\cite{KASTRUP} and references therein). In the present paper, we do
not aim at a discussion of the still interesting issue of
angle-operators and related problems. We will concentrate on an
important tool for a theoretical analysis of cyclic quantum systems,
the generalized coherent states.

The coherent states on the real line provide a quite useful tool for
a semiclassical analysis of one-dimensional systems. Their
generalization to field theory is the main tool for describing laser
light and other states of coherently superposed matter. This has
raised the question if one can find states with equally pleasant
properties for other systems, such as quantum mechanic on the circle
\cite{KASTRUP,KRP,KR2,KR}, or for other dynamical groups, such as
the states defined by Perelomov or Gilmore \cite{GIL90}.

In \cite{CCS}, a method was described to construct coherent states
for quantum mechanics on any compact, connected Lie group $G$. The
mathematical foundation for this was laid in \cite{HALL1, HALL2,
HALL3}. The procedure relies on the fact that the tangent bundle
$T^*G$, which can be identified with the phase space of a particle
moving on $G$, is diffeomorphic to the complexification $G^{\C}$
of $G$, which named these states "complexifier coherent states".
Their general properties have been exhibited in \cite{GCS1, GCS2,GCS3}.
For $G=SU(2)$, the complexifier coherent states are used
for the semiclassical analysis of Loop Quantum Gravity, and the
states for $G=U(1)$ and $G=U(1)^3$ are employed in the
semiclassical analysis of quantum cosmology \cite{ABL} and
linearized quantum gravity \cite{CCS}.

For the special case of $G=U(1)$, i.e.~quantum mechanics on the
circle, the complexifier coherent states reduce to the already
employed coherent states on the circle, which had either been
derived from a generalization of the harmonic oscillator coherent
states \cite{KRP, KR, GONZ}, or simply guessed \cite{CS86,KMW}. In
\cite{KASTRUP}, one can find an extensive description of these
states and their properties.

 In this article, we will shortly review the most important properties of the
complexifier coherent states for $G=U(1)$ and demonstrate that
these states can be used for a semiclassical analysis of quantum
mechanics on the circle, toward which we will pursue two
different ways.

First, we will analyze the Bargmann-Segal representation of states
on the circle, defined by the complexifier coherent states. In
particular, this enables one to study the dynamics of
one-dimensional periodic systems by the zeros of the Husimi
distribution. For quantum mechanical systems on the real line the
Husimi distribution is a well-developed tool to study, e.g., quantum
chaos \cite{KMW, NONNE}. In our work, we will show that the
complexifier coherent states on the circle allow for a similar
definition of the Husimi distribution of a state and that one needs
even less effort to reconstruct the state from its zeros than in the
case of quantum mechanics on the real line. This could serve as a
starting point of investigating quantum chaotic behavior of periodic
systems, such as happened for the driven rotor \cite{GKM}.

In the second part of this paper, we derive the semiclassical
propagator on a cylindrical phase space, which is the transition amplitude $\langle
z_F|\exp(-\rmi \hat H \t)|z_I\rangle$ between complexifier
coherent states. Semiclassical propagators have a long history
 and have been studied in detail, e.g.~for systems on
the real line, without or with spin (see, e.g., \cite{KECK,Pari03,Ribe04,Nova05,Ribe05,Pari06}
and references therein). They are a main ingredient for the
semiclassical analysis of quantum systems, and a powerful tool to
investigate the transition from quantum to classical behavior of
systems. The derivation of the semiclassical propagator on the
circle follows similar lines as the one for the propagator on the
real line \cite{KECK}. Finally, the propagator on the circle can be
represented as an infinite sum over propagators on the real line,
representing different winding numbers of
paths 
in $U(1)$, which nicely demonstrates the influence of the
global topological properties of phase space on the quantum dynamics.

\section{Quantum mechanics on the circle}\label{Ch:Kapitel1}

\noindent The phase space for a classical particle moving on a
circle $U(1)$ is given by $T^*(S^1)\simeq S^1\times \R$. Since the
angle $\phi$ is not periodic, it is no phase space function. So
one has to work with $\exp(\rmi\phi)$, of which one can compute
$\phi$ only modulo $2\pi$. Together with the canonical momentum
$p$, which has the dimension of angular momentum, the Poisson
bracket:
\begin{eqnarray}
\big\{\,f\,,\,g\,\big\}\;=\;\frac{\del f}{\del\phi}\frac{\del
g}{\del p}-\frac{\del g}{\del\phi}\frac{\del f}{\del p}
\end{eqnarray}

\noindent implies:
\begin{eqnarray}\label{Gl:KommutatoraufKreis}
\big\{\,\exp(\rmi\phi)\,,\,p\,\big\}\;=\;\rmi\,\exp(\rmi\phi).
\end{eqnarray}

\noindent Note that, although a function on phase space,
$\exp(\rmi\phi)$ is not an observable, since it is not real-valued.

 Quantization is achieved by replacing observables by
selfadjoint operators and Poisson brackets by commutators divided
by $\rmi\hbar$. Since $\exp(\rmi\phi)$ is not a real-valued
function, but takes values in $U(1)$, we will replace it by a
unitary operator $\exp(\rmi\hat\phi)$. The quantization of
(\ref{Gl:KommutatoraufKreis}) then yields:

\begin{eqnarray}\label{Gl:KommutatoraufKreisQuantum}
\big[\,\exp(\rmi\hat\phi)\,,\,\hat{p}\,\big]\;=\;\rmi\hbar\,(\rmi
\exp(\rmi\hat\phi))\;=\;-\hbar \,\exp(\rmi\hat\phi).
\end{eqnarray}

\noindent The quantization of (\ref{Gl:KommutatoraufKreis}) is
nontrivial, since there are infinitely many unitarily inequivalent
representations of (\ref{Gl:KommutatoraufKreisQuantum})
\cite{KASTRUP, KRP}. The different representations live all on the
Hilbert space
\begin{eqnarray}
\H\,=\,L^2[0,2\pi]
\end{eqnarray}

\noindent with the inner product

\begin{eqnarray}\label{Gl:SkalarproduktaufdemKreis}
\langle\,\psi\,,\,\varphi\,\rangle\;=\;\int_0^{2\pi}\,\frac{\rmd\phi}{2\pi}\;\overline{\psi(\phi)}\,\varphi(\phi),
\end{eqnarray}

\noindent where the operator $\exp(\rmi\hat\phi)$ acts as
multiplication
\begin{eqnarray}\label{Gl:ActionofHolonomy}
\left(\exp(\rmi\hat\phi)\psi\right)(\phi)\;:=\;\exp(\rmi\phi)\psi(\phi),
\end{eqnarray}

\noindent and $\hat p$ as differentiation
\begin{eqnarray}
(\hat p
\psi)(\phi)\;=\;\frac{\hbar}{\rmi}\,\frac{\rmd\psi}{\rmd\phi}(\phi).
\end{eqnarray}

\noindent While $\exp(\rmi\hat\phi)$ is a unitary operator, hence
bounded by one, it is defined everywhere, whereas $\hat p$ is
unbounded, and one has to worry about domains of definition. In
particular, the inequivalent representations of
(\ref{Gl:KommutatoraufKreisQuantum}) differ by the dense domain of
definition of $\hat p$. They are labeled by a real parameter
$0\leq\d<1$. Fixing $\d$, one defines a basis of $\H$ by
\begin{eqnarray}\label{Gl:Eigenvectors}
|n\rangle_{\d}\:=\;\phi\mapsto
\exp\big({\rmi(n+\d)\phi}\big),\quad n\in\Z.
\end{eqnarray}

\noindent The domain of definition for $\hat p$ in the
representation labeled by $\d$ is the given by the following set
of linear combinations of $|n\rangle_{\d}$:
\begin{eqnarray}\label{Gl:DomainOfDefinition}
\D^{\d}_{\hat
p}:=\left\{\sum_{n\in\Z}c_n|n\rangle_{\d}\;\Bigg|\;\sum_n
n^2|c_n|^2<\infty\right\}.
\end{eqnarray}

\noindent Note that in each representation $\d$, $\hat p$ has the
basis (\ref{Gl:Eigenvectors}) as a complete set of eigenvectors:
\begin{eqnarray}\label{Gl:EigenvaluesOfP}
\hat p|n\rangle_{\d}\;=\;(n+\d)\hbar|n\rangle_{\d}.
\end{eqnarray}

\noindent Equation (\ref{Gl:EigenvaluesOfP}) shows that
representations for different $\d$ are in fact unitarily
inequivalent. The parameter $\d$ determines the fraction of $2\pi$
the phase of a particle acquires, if it runs once round the circle.
In periodic crystals, the parameter $\d$ can be seen as the
quasimomentum of the Bloch waves \cite{KASTRUP}. The meaning of the
parameter $\d$ will be illustrated later in more detail, when we
derive the semiclassical propagator on the circle, which heavily
depends on $\d$, since it keeps track of different paths with
different winding number, where the relative phase is crucial.

In \cite{KRP}, it was reasoned that the cases $\d=0$ and $\d=1/2$
are the most important ones, since in these representations the
systems are invariant under time-reversal symmetry. In these
representations one cannot distinguish between a wave running
clockwise or anti-clockwise around the circle, since the first one
acquires a phase of $e^{2\pi \rmi \d}$ and the other one a phase of
$e^{-2\pi \rmi \d}$, which is not the same for $\d\neq
0,\,\frac{1}{2}$. Still, one could imagine systems where nontrivial
phase shifts occur due to complicated interactions, which explicitly
distinguish between clockwise or anti-clockwise moving waves. In
particular, the topic of fractional optical angular momentum (OAM)
is discussed in \cite{KASTRUP}. So in what follows, we will not fix
$\d\in[0,1)$, to keep the results as general as possible.

\section{Coherent States}

\subsection{Definition}

\noindent On the circle, as well as on any other  compact,
connected Lie group, one can define the "complexifier coherent
states" \cite{CCS}, which for the case of the circle are different
for every $\d$. They are furthermore labeled by a squeezing
parameter $s>0$ and a complex number $z=\phi+\rmi{p}/{\hbar}$:

\begin{eqnarray}\label{Gl:CoherentStates}
|z\rangle_{\d}\;&=\;\sum_{n\in\Z}\exp\left({-(n+\d)^2\frac{s^2}{2}}\;{-\;\rmi(n+\d)z}\right)|n\rangle_{\d}\nonumber\\[5pt]
\;&=\;\sum_{n\in\Z}\exp\left({-(n+\d)^2\frac{s^2}{2}}\;+\;{(n+\d)\frac{p}{\hbar}}\;{-\;\rmi(n+\d)\phi}\right)
|n\rangle_{\d}.
\end{eqnarray}

\noindent Note that the map $z\mapsto|z\rangle_{\d}$ is periodic
in the sense that
\begin{equation}
|z+2\pi\rangle_{\d}\;=\;\rme^{-2\pi \rmi \d}\,|z\rangle_{\d}.
\end{equation}

\noindent The states (\ref{Gl:CoherentStates}) are not normalized,
but instead we have
\begin{eqnarray}\label{Gl:CoherentNorm}
\|z_{\d}\|^2\;:=\;{}_{\d}\langle
z|z\rangle_{\d}\,=\,\sum_{n\in\Z}\exp\left({-(n+\d)^2s^2\;+\;2(n+\d)\frac{p}{\hbar}}\right).
\end{eqnarray}

\noindent The infinite sum in (\ref{Gl:CoherentNorm}) can be
brought into a nicer form by the so-called "Poisson summation
formula"  (see e.g.~\cite{JF98}), given by

\begin{eqnarray}\label{Gl:PoissonResummationsFormel}
\sum_{n\in\Z}f(an)\;=\;\frac{1}{a}\sum_{n\in\Z}\tilde{f}\left(\frac{2\pi
n}{a}\right),
\end{eqnarray}

\noindent where $\tilde{f}$ is the Fourier transform of $f$:
\begin{eqnarray}
\tilde{f}(k)\,=\,\int_{\R}dx\,f(x)\,e^{-\rmi kx}.
\end{eqnarray}

\noindent Choosing
$f(x)=\exp\left({-x^2s^2\,+\,2x{p}/{\hbar}}\right)$, with
\begin{eqnarray}
\tilde{f}(k)\,=\,\sqrt{\frac{\pi}{s^2}}\,\exp\left({\frac{p^2}{s^2\hbar^2}}\;-\;\frac{k^2-4\rmi
k p/\hbar}{4s^2}\right),
\end{eqnarray}

\noindent equation (\ref{Gl:CoherentNorm}) can be rewritten as

\begin{eqnarray}
\fl\|z_{\d}\|^2\,&=\,\sqrt\frac{
\pi}{s^2}\;\exp\left({\frac{p^2}{s^2\hbar^2}}\right)\,\sum_{n\in\Z}\exp\left({-2\pi
n\d
\rmi}\;{-\;\frac{\pi^2n^2\;-\;\rmi\pi np/\hbar}{s^2}}\right)\nonumber\\[5pt]
\fl&=\,\sqrt\frac{
\pi}{s^2}\;\exp\left({\frac{p^2}{s^2\hbar^2}}\right)\,\left[1+\sum_{n\neq
0}\left({-2\pi n\d \rmi\;-\;\frac{\pi^2n^2-\rmi\pi
np/\hbar}{s^2}}\right)\right].
\end{eqnarray}

\noindent One can easily show the inequality

\begin{eqnarray}\label{Gl:ShittyOverlap}
\left|\sum_{n\neq 0}\exp\left(-2\pi n\d
\rmi\,-\,\frac{\pi^2n^2-\rmi\pi
np/\hbar}{s^2}\right)\right|\;\leq\;2\frac{e^{-\frac{\pi^2}{s^2}}}{1-e^{-\frac{\pi^2}{s^2}}},
\end{eqnarray}

\noindent which tends to zero faster than any power of $s$.
Therefore we denote it as $O(s^{\infty})$ and write

\begin{eqnarray}\label{Gl:CoherentBetterNorm}
\|z_{\d}\|^2\;=\;\sqrt\frac{
\pi}{s^2}\;\exp\left({\frac{p^2}{s^2\hbar^2}}\right)\,\left(1+O(s^{\infty})\right).
\end{eqnarray}

\noindent The inner product between two coherent states
$|z\rangle_{\d}$ and $|w\rangle_{\d}$ can be obtained by a similar
calculation, using again the Poisson summation formula:
\begin{eqnarray}\label{Gl:CoherentOverlap}
\fl{}_{\d}\langle z' | z
\rangle_{\d}\;&=\;\sqrt{\frac{\pi}{s^2}}\,\sum_{n\in\Z}\,\exp\left[{2\pi \rmi n \d}
\;-\;\left(n\pi-\frac{\overline{z}'-z}{2}\right)^2\frac{1}{s^2}\right]\nonumber\\[5pt]\nonumber
\fl\;&=\;\sqrt{\frac{\pi}{s^2}}\,\exp\left[\left(\frac{p+p'}{2s\hbar}\right)^2\right]
\,\sum_{n\in\Z}\,\exp\left[2\pi \rmi n \d\;
-\;\left(\frac{\phi'-\phi-2\pi
n}{2s}\right)^2\right.\\[5pt]
\fl&\qquad\qquad\qquad\left.\;+\;2\rmi\left(\frac{\phi'-\phi-2\pi
n}{2s}\right)\frac{p'+p}{2s\hbar}\right].
\end{eqnarray}

\noindent Note that (\ref{Gl:CoherentOverlap}), as well as the
formula for the coherent states itself (\ref{Gl:CoherentStates}) can
be written (see e.g. \cite{KASTRUP, KRP}) in terms of the Jacobian
theta function of third kind
\begin{eqnarray}\label{Gl:ThetaFunction}
\vartheta_3(q,\,\t)\;=\;\sum_{n\in\Z}\,\exp\left(\pi\rmi
n^2\t\,+\,2\pi\rmi n z\right).
\end{eqnarray}

\noindent Although this form is quite general and, since the
properties of the theta-functions are well-known \cite{APOSTOL},
(\ref{Gl:ThetaFunction}) shows a lot about the structure of the
coherent states, we will keep the explicit form
(\ref{Gl:CoherentOverlap}), which will prove to be more convenient
for the analytic treatment of the expressions in the sections
\ref{Ch:Hadamard} and \ref{Ch:Propagator}.\\

From (\ref{Gl:CoherentOverlap}) this and
(\ref{Gl:CoherentBetterNorm}) we obtain the overlap of two coherent
states as:
\begin{eqnarray}\nonumber
\fl\frac{{}_{\d}\langle z' | z
\rangle_{\d}}{\|z'_{\d}\|\,\|z_{\d}\|}\;=\;&\exp\left[-\left(\frac{p'-p}{2s\hbar}\right)^2\right]
\sum_{n\in\Z}\;\exp\left[2\pi \rmi n \d\;-\;\left(\frac{\phi'-\phi-2\pi n}{2s}
\right)^2\right.\\[5pt]\label{Gl:CoherentBetterOverlap}
\fl&\left.\;+\;2\rmi\left(\frac{\phi'-\phi-2\pi
n}{2s}\right)\frac{p'+p}{2s\hbar}\right](1+O(s^{\infty})).
\end{eqnarray}

\noindent From formula (\ref{Gl:CoherentBetterOverlap}) one can
see that coherent states $|z\rangle_{\d}$ and $|z'\rangle_{\d}$
with $z\neq z'$, i.e.~states labeled by different points
$z=\phi+\rmi p/\hbar$ and $z'=\phi'+\rmi p'/\hbar$, have an
overlap that tends to zero faster than any power of $s$, so the
overlap function (\ref{Gl:CoherentBetterOverlap}) is peaked at
$z=z'$, the peak becoming sharper as $s\to 0$. If $\phi'\neq
\pm\pi$ and $s\ll \pi$, then all terms in the infinite sum
(\ref{Gl:CoherentBetterOverlap}) are of order $O(s^{\infty})$, so
we get
\begin{eqnarray}
\fl\frac{{}_{\d}\langle z' | z
\rangle_{\d}}{\|z'_{\d}\|\,\|z_{\d}\|}\;=\;\exp\left[-\left(\frac{p'-p}{2s\hbar}\right)^2\,
-\,\left(\frac{\phi'-\phi}{2s} \right)^2\;\right.\\\nonumber
\qquad\qquad\left.+\;2\rmi\left(\frac{\phi'-\phi}{2s}\right)\frac{p'+p}{2s\hbar}\right](1+O(s^{\infty})).
\end{eqnarray}

\noindent In particular, the overlap for fixed $z'$ is -- up to
small corrections in $s$ -- a Gaussian in the complex $z$-plane
centered at $z=z'$, with width $s$.\\

\noindent Note that (\ref{Gl:Eigenvectors}) defines a
$\phi$-representation via
$\langle\phi|n\rangle_{\d}=\exp[-\rmi(n+\d)\phi]$. Using
(\ref{Gl:PoissonResummationsFormel}), one can show that in this
representation the coherent states are infinite superpositions of
Gaussian wavepackets with a width $s$, each translated by
$\phi\to\phi+2\pi n$, ensuring that the resulting function in
$2\pi$-periodic in $\phi$. This shows that the complexifier
coherent states on the circle are the periodically continued
harmonic oscillator coherent states, which have been used earlier
\cite{CS86,GKM}. As long as the width $s$ of these wavepackets is
much smaller than the period $2\pi$, the different Gaussians
interfere little with each other, and one can restrict oneself to
one Gaussian in calculations. But if the spreading $s$ exceeds,
say, $\pi$, one has to take the infinite mutual interference of
the Gaussians into account, which makes calculations quite
difficult.\\

Since the parameter $s$ measures the spreading of the wavefunction
$\psi_z(\phi)\,=\, \langle\phi|z\rangle_{\d}$ compared to the
circumference of the circle on which the system propagates, the complexifier coherent states only describe systems
being "close to classical point particles" if $s$ is small:
$s\ll\pi$. This feature is quite natural, since, if a particle
moves on a circle with a de Broglie wavelength the same order of
magnitude as the circumference of the circle, one cannot expect
this particle to behave classically. The wavefunction of the
particle will interfere with itself "around the circle", which is
not possible in classical mechanics. This is why the states are
called "coherent" rather than "semiclassical states", and is
simply due to the compact topology of configuration space. The
limit in which quantum mechanics on the real line is recovered is
then performed as $s\to 0$.

\subsection{Properties}

\noindent The complexifier coherent states (\ref{Gl:CoherentStates})
have a number of properties they share with ordinary harmonic
oscillator coherent states, which have been exhibited in
\cite{KASTRUP, KRP, GCS1, GCS2, GCS3} and are listed for the sake of
completeness.

\begin{itemize}
\item
\noindent{Reproduction of classical values:}\\

\noindent The expectation values of certain operators in the
coherent states labeled by $z=\phi+\rmi p/\hbar$ agree -- up to
small corrections in $s$ -- with the value of the corresponding
classical phase space functions, evaluated at the points
$(\phi,\,s^{-2}p)$.  This will be demonstrated with the basic
operators $\exp(\rmi\hat\phi)$ and $\hat p$. We start with
$\exp(\rmi\hat\phi)$. For this we remember
(\ref{Gl:ActionofHolonomy}) and (\ref{Gl:Eigenvectors}), in
particular
\begin{eqnarray}\label{Gl:ActionofHolonomyInBasis}
\exp(\rmi\hat\phi)|n\rangle_{\d}\;=\;|n+1\rangle_{\d}.
\end{eqnarray}

\noindent With  $z=\phi+{\rmi}p/{\hbar}$ and the definition of the
coherent states (\ref{Gl:CoherentStates}) we get
\begin{eqnarray}
\fl{}_{\d}\langle z|\,\exp(\rmi\hat\phi)\,| z
\rangle_{\d}\;&=\;\sum_{n\in\Z}\exp\Bigg({-(n+\d)^2\frac{s^2}{2}-(n-1+\d)^2\frac{s^2}{2}}\nonumber\\[5pt]
&\qquad\qquad\;+\;{(2n+2\d-1) \frac{p}{\hbar}}\;+\;\rmi\phi\Bigg).
\end{eqnarray}

\noindent With the help of the Poisson summation formula
(\ref{Gl:PoissonResummationsFormel}), we obtain after some
straightforward computation:

\begin{eqnarray}
\fl{}_{\d}\langle z|\,\exp(\rmi\hat\phi)\,| z
\rangle_{\d}\;=\;\sqrt\frac{\pi}{s^2}\,\,\exp\left(\rmi\phi\,-\,\frac{s^2}{4}
\;+\;\frac{p^2}{s^2\hbar^2}\right)\\[5pt]\nonumber
\qquad\times\;\sum_{n\in\Z}\;\exp\left(2\pi n\d
\rmi\;-\;\frac{\pi^2n^2+\rmi n\pi p/\hbar-\rmi\pi n}{s^2}\right).
\end{eqnarray}

\noindent By using (\ref{Gl:CoherentBetterNorm}) and again
estimating all terms with $n\neq 0$:
\begin{eqnarray}
\sum_{n\neq 0}\left(2\pi n\d \rmi\;-\;\frac{\pi^2n^2+\rmi n\pi
p/\hbar-\rmi\pi n}{s^2}\right)\;=\;O(s^{\infty}),
\end{eqnarray}

\noindent we arrive at the result

\begin{eqnarray}\label{Gl:PhaseExpValue}
\frac{{}_{\d}\langle z|\,\exp(\rmi\hat\phi)\,| z
\rangle_{\d}}{{}_{\d}\langle
z|z\rangle_{\d}}\;&=&\;\,\exp\left({\rmi\phi-\frac{s^2}{4}}\right)\,(1+O(s^{\infty}))\nonumber\\[5pt]
&=&\;\,\exp\left({\rmi\phi}\right)\,(1+O(s^2))
\end{eqnarray}

\noindent So, up to order $O(s^2)$, the expectation value of
$\exp(\rmi\hat\phi)$ in coherent states labeled by the complex
number $z=\phi+{\rmi}p/{\hbar}$ agrees with the value of the
classical phase space function $\exp(\rmi\phi)$ at this point. By
a similar calculation, one can even show that the expectation
value of the relative phase between two particles in coherent
states labeled by $z,\,z'$ is $\exp[\rmi(\phi-\phi')]$, up to
$O(s^{\infty})$-corrections.\\

\noindent We continue with $\hat p$. From
(\ref{Gl:EigenvaluesOfP}) and (\ref{Gl:PoissonResummationsFormel})
we get

\begin{eqnarray}
\fl{}_{\d}\langle z|\,\hat p\,| z
\rangle_{\d}\;&=\;\hbar\sum_{n\in\Z}(n+\d)\;\exp\left(-(n+\d)^2s^2\;+\;2(n+\d)\frac{p}{\hbar}\;\right)\;\\[5pt]\nonumber
\fl&=\;\sqrt\frac{\pi}{s^2}\,\exp\left(\frac{p^2}{s^2\hbar^2}\right)\,\sum_{n\in\Z}\left(\frac{p}{s^2}-\pi
n\hbar\right)\exp\left(2\pi n\d \rmi\;-\;\frac{\pi^2n^2+\rmi p
n/\hbar}{s^2}\right) .
\end{eqnarray}

\noindent With this and (\ref{Gl:CoherentBetterNorm}) we obtain
the result
\begin{eqnarray}
\frac{{}_{\d}\langle z|\,\hat p\,| z \rangle_{\d}}{{}_{\d}\langle
z|z\rangle_{\d}}\;=\;\frac{p}{s^2}(1+O(s^{\infty})).
\end{eqnarray}

\noindent So the coherent state labeled by the complex number
$z=\phi+{\rmi}p/{\hbar}$ is peaked around the phase space
point $(\phi, s^{-2}p)$.\\

Note that the expectation values of the operators
$\exp(\rmi\hat\phi)$ and $\hat p$ give the classical values
$\exp(\rmi\phi)$ and $s^{-2}p$, respectively, only up to corrections in $s$.
If $s\to 0$ and $s^{-2}p$ is held fixed, the corrections go to zero, which
corresponds to the radius of the circle going to infinity, suppressing the
self-interference of the wavefunction. Quantum mechanics on the real
line is recovered in that limit. In particular, the expectation
values of the operators $\hat X$ and $\hat P$ in the coherent states
$|z\rangle=|q+\rmi p\rangle$ on the real line reproduce the classical
values $q$ and $p$ exactly.\\

\item\noindent{Resolution of the identity:}\\

\noindent Apart from reproducing classical phase space function
values, the coherent states form an overcomplete system of vectors
(see section \ref{Ch:Hadamard}). With $z=\phi+\rmi p/{\hbar}$, we
get

\begin{eqnarray}\label{Gl:CoherernOvercomplete}
\frac{1}{\sqrt{\pi} s\hbar}\int_{\R}\rmd
p\int_{-\pi}^{\pi}\frac{\rmd\phi}{2\pi}\,\exp\left(-\frac{p^2}{s^2\hbar^2}\right)
\,|z\rangle_{\d}{}_{\d}\langle z|\;=\;\mathbbm{1},
\end{eqnarray}

\noindent which can be shown easily by calculating the action of
the left hand side of (\ref{Gl:CoherernOvercomplete}) on
a basis vector $|n\rangle_{\d}$.\\

\item\noindent{Ladder operator eigenstates:}\\

\noindent Both operators $\exp\left({-s^2\hat
p^2/2\hbar^2}\right)$ and $\exp({\rmi\hat\phi})$ are bounded, so
the operator
\begin{eqnarray}\label{Gl:LadderOperators2}
\hat g\;=\;\exp\left({-s^2\frac{\hat
p^2}{2\hbar^2}}\right)\,\exp({\rmi\hat\phi})\,\exp\left({s^2\frac{\hat
p^2}{2\hbar^2}}\right)
\end{eqnarray}

\noindent is well-defined on the domain of definition of $\hat p$
(\ref{Gl:DomainOfDefinition}).  We have
\begin{eqnarray}\label{Gl:LadderOperators}
\hat
g|n\rangle_{\d}\;=\;\exp\left[{\frac{s^2}{2}\Big((n+\d)^2-(n+\d+1)^2\Big)}\right]|n+1\rangle_{\d}.
\end{eqnarray}

\noindent With this one obtains
\begin{eqnarray}
\hat g
|z\rangle_{\d}\;&=\;\sum_{n\in\Z}\exp\left({-(n+\d)^2\frac{s^2}{2}}\;-\;{\rmi(n+\d)z}\right)\,\hat
g|n\rangle_{\d}\nonumber\\[5pt]\nonumber
&=\;\sum_{n\in\Z}\exp\left({-(n+1+\d)^2\frac{s^2}{2}}\;-\;{\rmi(n+\d)z}\right)\,|n+1\rangle_{\d}\\[5pt]
&=\;\rme^{\rmi z}\;|z\rangle_{\d}.
\end{eqnarray}

\noindent So the coherent states labeled by $z$ are eigenvectors
of $\hat g$ with eigenvalue $\rme^{\rmi z}$. With
(\ref{Gl:LadderOperators}) one can see that $\hat g$ and $\hat
g^{\dag}$ are in fact ladder operators, although their commutator
is not proportional to unity. Note that, by formally using the
Baker-Campbell-Hausdorff formula, one can bring $\hat g$ into the
form of $\hat g=\exp{\rmi(\hat\phi+{\rmi}\hat p/{\hbar})}$.\\

\item\noindent{Minimal uncertainty relationship:}\\

\noindent As demonstrated in \cite{CCS} or \cite{KRP}, the fact
that the coherent states are eigenvectors of $\hat g$ immediately
implies that the coherent states saturate the Heisenberg
inequality for the operators
\begin{eqnarray}
\hat Q\;=\;\frac{\hat g+\hat g^{\dag}}{2}\;,\qquad \hat
P\;=\;\frac{\hat g-\hat g^{\dag}}{2i},
\end{eqnarray}

\noindent that is
\begin{eqnarray}
\Delta_z\hat Q\,\Delta_z\hat
P\;=\;\frac{1}{2}\;\Big\langle\left[\hat Q,\,\hat
P\right]\Big\rangle_z,
\end{eqnarray}

\noindent where $\langle\cdot\rangle_{z}$ and $\Delta_z$ denote
 expectation value and
standard deviation in $|z\rangle_{\d}$ respectively.

\end{itemize}

\section{Bargmann-Segal representation and Hadamard
decomposition}\label{Ch:Hadamard}

\noindent In quantum mechanics on the real line, the harmonic
oscillator coherent states $|z\rangle$ provide the Bargmann-Segal
representation $\psi(z)=\langle\psi|z\rangle$ for Hilbert space
states $\psi$. From this one can construct the Husimi-distribution
\cite{KMW, VO06}
\begin{eqnarray}
 \rho_{\rm Husimi}(q,p)\;=\;\exp\left(-p^2-q^2\right)\big|\psi(q+\rmi p)\big|^2.
\end{eqnarray}

\noindent This phase-space density provides a way to analyze
dynamical properties of systems, in particular the behavior of
classically chaotic quantum systems \cite{KMW, WK97}. It is
possible to reconstruct the state $\psi$ from the zeros of its Husimi
distribution and the specification of three complex numbers
$C_0,\,C_1,\,C_2$ \cite{WK97}, which have to be derived from
$\rho_{\rm Husimi}$.

 In the following we will describe the analogous
construction for the complexifier coherent states on the circle
and show that the condition of periodicity in one phase space
variable poses significant limitations to the form of the
Bargmann-Segal representation. In particular, the zeros of the
Husimi distribution of a state determine this state apart from
normalization and specification of an integer, which can be
computed from $\rho_{\rm Husimi}$.

 The Bargmann-Segal representation of a state
$\psi$ on the circle, provided by the complexifier coherent
states,  is given by the function
\begin{eqnarray}\label{Gl:Bargmann-Segal}
\fl\psi_{\d}(z)\;&=&\;\langle\psi|z\rangle_{\d}\;=\;\sum_{n\in\Z}\exp\left({-(n+\d)^2\frac{s^2}{2}}\;-\;{\rmi(n+\d)z}\right)\langle\psi|n\rangle_{\d}.
\end{eqnarray}

\noindent The map (\ref{Gl:Bargmann-Segal}) is entire holomorphic.
Note that (\ref{Gl:CoherernOvercomplete}) guarantees that
\begin{eqnarray}
\frac{1}{\sqrt{\pi }s\hbar}\int_{\R}\rmd
p\int_{-\pi}^{\pi}\frac{\rmd\phi}{2\pi}\,\exp\left({-\frac{p^2}{s^2\hbar^2}}\right)
\;|\psi_{\d}(z)|^2\;=\;\langle\psi|\psi\rangle,
\end{eqnarray}

\noindent which shows that the map between the vector
$|\psi\rangle$ and the function $z\mapsto\psi_{\d}(z)$ is an
anti-unitarity between the Hilbert space $L^2[0,2\pi]$ and the
Hilbert space $\H L^2(S^1\times\R, d\mu)$. With the latter we mean
the Hilbert space of periodic, holomorphic functions of
$z=\phi+{\rmi}p/{\hbar}$ that are square-integrable with respect
to the measure
\begin{eqnarray}
\rmd\mu(z)\;=\;\frac{1}{\sqrt{\pi}s\hbar}\;\rmd
p\,\frac{\rmd\phi}{2\pi}\;\exp\left({-\frac{p^2}{s^2\hbar^2}}\right).
\end{eqnarray}

\noindent That (\ref{Gl:Bargmann-Segal}) is holomorphic has an
important corollary: The set of coherent states $|z\rangle_{\d}$
spans the Hilbert space, hence it is complete. It is even
overcomplete in the sense that smaller subsets of it also span the
Hilbert space. Let $\{z_n\}_{n\in\N}$ be a sequence of distinct
complex numbers converging to $z\in\C$. Then the set of coherent
states $\{|z\rangle_{\d}\}_{n\in\N}$ is complete. This is quite easy
to show by noting that, if $\psi$ is orthogonal to the span of the
$|z_n\rangle_{\d}$, then $\psi_{\d}(z_n)=0$ for all $n$. But since
$\psi_{\d}(z)$ is holomorphic and the $z_n$ converge, then
$\psi_{\d}\equiv 0$ by the identity theorem for holomorphic
functions. So the linear span of the $|z_n\rangle_{\d}$ is dense
in $L^2[0,2\pi]$.\\

Just as in the case of harmonic oscillator coherent states, the
holomorphic functions $\psi_{\d}(z)$ are entire and of order at
most two:

\begin{Proposition} Let $|\psi\rangle\in\H$. Then there exist
constants $A, B>0$, such that
\begin{eqnarray}
|\psi_{\d}(z)|\;=\;|\langle\psi|z\rangle_{\d}|\;\leq\;A\exp\left({B|z|^2}\right).
\end{eqnarray}
\end{Proposition}

\noindent\textbf{Proof:} With $z=\phi+\rmi p/{\hbar}$ we get,
using the Poisson summation formula
(\ref{Gl:PoissonResummationsFormel}):
\begin{eqnarray}\nonumber
\fl|\psi_{\d}(z)|\,\exp\left({-\frac{|z|^2}{2s^2}}\right)&\;\leq\;|\psi_{\d}(z)|\,\exp\left({-\frac{p^2}{2s^2\hbar^2}}\right)\\[5pt]\
&\;=\;\left|\sum_{n\in\Z}\exp\left({-(n+\d)^2\frac{s^2}{2}\;-\;\rmi(n+\d)z\;-\;\frac{p^2}{2s^2\hbar^2}}\right)
\langle\psi|n\rangle_{\d}\right|\nonumber\\[5pt]\nonumber
&\;\leq\;\sum_{n\in\Z}\exp\left(-(n+\d)^2\frac{s^2}{2}\;+\;(n+\d)\frac{p}{\hbar}\;-\;\frac{p^2}{2s^2\hbar^2}\right)
\left|\langle\psi|n\rangle_{\d}\right|\\[5pt]\nonumber
&\;\leq\;\sum_{n\in\Z}\exp\left[{-\left(n+\d-\frac{p}{s^2\hbar}\right)^2\frac{s^2}{2}}\right]
\psi_{\rm max}\\[5pt]\nonumber
&\;=\;\psi_{\rm
max}\,\sqrt\frac{2\pi}{s^2}\,\sum_{n\in\Z}\,\rme^{2\pi \rmi
n\d}\,\exp\left(-\frac{2\pi^2n^2+2\pi n \rmi p/\hbar}{s^2}\right)\\[5pt]
&\;\leq\;\psi_{\rm
max}\,\sqrt\frac{2\pi}{s^2}\,\sum_{n\in\Z}\,\exp\left(-\frac{2\pi^2n^2}{s^2}\right),
\end{eqnarray}

\noindent with $\psi_{\rm
max}=\max\left\{|\langle\psi|n\rangle_{\d}|\;,n\in\Z\right\}$, and
therefore
\begin{eqnarray}
|\psi(z)|\;\leq\;\psi_{\rm
max}\,\sqrt\frac{2\pi}{s^2}\,\sum_{n\in\Z}\,
\exp\left(-\frac{2\pi^2n^2}{s^2}\right) \;\exp\left(
{\frac{|z|^2}{2s^2}}\right),
\end{eqnarray}

\noindent what was to be shown.

 In fact, two is the best estimate one can give for the
order of the holomorphic functions $\psi_{\d}(z)$. Although there
are functions that are of order one (in particular $z\mapsto
{}_{\d}\langle n | z\rangle_{\d}$), there are also examples of
functions of order two (for instance the one belonging to
$|\psi\rangle\,=\,\sum_{n\neq 0}n^{-2}\,|n\rangle_{\d}$).\\

 Each entire holomorphic function can be expressed in
terms of its zeros. This representation is called the Hadamard
decomposition (see, e.g., \cite{RC95}). In particular, for
$\psi(z)$ being entire holomorphic and of order at most two, there
are constants $m\in\N$, $C_0,\,C_1,\,C_2\in\C$ such that

\begin{eqnarray}\label{Gl:Hadamard}
\psi(z)\;=\;z^m\,\rme^{C_0+C_1z+C_2z^2}\;\prod_{n}\left(1-\frac{z}{z_n}\right)
\,\exp\left[{\frac{z}{z_n}+\frac{1}{2}\left(\frac{z}{z_n}\right)^2}\right],
\end{eqnarray}

\noindent where the $\{z_n\}$ are the zeros of $\psi(z)$.

 The
entire holomorphic functions $\psi_{\d}(z)$
(\ref{Gl:Bargmann-Segal}) have the following properties:
\begin{eqnarray}\label{Gl:Hadamard:Quasiperiodicity}
\psi_{\d}(z+2\pi)\;=\;\rme^{-2\pi \rmi\d}\,\psi_{\d}(z).
\end{eqnarray}

\noindent  Therefore their number of zeros can only be zero or
infinity. On the other hand, one only has to know the zeros in the
strip $[0,2\pi)\times \rmi\R\subset\C$ to know all zeros of
$\psi(z)$.

Since $\psi_{\d}$ is of order at most two (which relates to the
growth behavior), the sequence of its zeros $\{z_n\}_{n\in\N}$ is
also of order at most two. For sequences this means that the
number of sequence-members contained in a circle of radius $R$
goes not faster than $R^2$, as $R$ grows large \cite{RC95}. It
follows that the sequence of zeros lying inside the strip
$[0,2\pi)\times \rmi\R\subset\C$
is of order at most one, which will be needed later.\\

Periodicity now restricts the possible values of $C_1,\,C_2$ in
(\ref{Gl:Hadamard}), as the following proposition shows:

\begin{Proposition}
Let $\psi_{\d}(z)$ be entire holomorphic in $z$, of order at most
two and $\psi_{\d}(z+2\pi)=\rme^{-2\pi \rmi\d}\psi_{\d}(z)$ . Then
there are constants $C\in\C$ and $m,\,l\in\Z$ such that
\begin{eqnarray}\label{Gl:HadamardPeriodic}
\fl
\psi_{\d}(z)\;=\;\rme^{C+i(l-\d)z}\;\left[\,\sin\frac{z}{2}\,\exp\left({-\rmi\frac{z}{2}}\right)\,\right]^m
\prod_{k}\;\left[\,\frac{\sin\frac{z-a_k}{2}}{\sin\frac{-a_k}{2}}\,\exp\left({-\n_k\rmi\frac{z}{2}}\right)\,\right],
\end{eqnarray}

\noindent where the $a_k$ are the zeros in the strip
$[0,2\pi)\times \rmi\R\subset\C$, apart from $0$, and
$\n_k\,:=\,{\rm sgn}\;{\rm Im}\,a_k$ the sign of the imaginary
part of $a_k$.
\end{Proposition}

\noindent\textbf{Proof:} We start with (\ref{Gl:Hadamard}) and
note that the zeros are $\{z_k+2\pi n\,|\,n\in\Z\}$, because of
(\ref{Gl:Hadamard:Quasiperiodicity}). Then we divert the possible
$m$-fold zeros at $2\pi n$:
\begin{eqnarray}\nonumber
\fl \psi_{\d}(z)\;=\;\left[z\prod_{n\neq 0}\left(1-\frac{z}{2\pi
n}\right)\exp\left({\frac{z}{2\pi
n}+\frac{1}{2}\left(\frac{z}{2\pi
n}\right)^2}\right)\right]^m\\[5pt]\nonumber
\fl\quad\qquad\;\times\;\rme^{C_0+C_1z+C_2z^2}\;\prod_{k}\prod_{n\in\Z}\left(1-\frac{z}{a_k+2\pi
n}\right)\exp\left[{\frac{z}{a_k+2\pi
 n}+\frac{1}{2}\left(\frac{z}{a_k+2\pi
n}\right)^2}\right]\\[5pt]\nonumber
\fl\qquad\;=\;\left[z\prod_{n=1}^{\infty}\left(1-\frac{z^2}{4\pi^2
n^2}\right)\exp\left({\frac{1}{4\pi^2
n^2}z^2}\right)\right]^m\;\rme^{C_0+C_1z+C_2z^2}\\[5pt]\label{Gl:Gleichung1}
\fl\quad\qquad\;\times\;\prod_{k}\Bigg[\left(1-\frac{z}{a_k}\right)\;\prod_{n=1}^{\infty}\left(1-\frac{z}{a_k+2\pi
n}\right)\left(1-\frac{z}{a_k-2\pi
n}\right)\nonumber\\[5pt]
\qquad\qquad\times\;\exp\left({\mathfrak
C_1(a_k)z+\frac{1}{2}\mathfrak C_2(a_k)z^2}\right)\Bigg]
\end{eqnarray}

\noindent Here the $a_k$ are the zeros inside the strip
$[0,2\pi)\times \rmi\R$ apart from $0$, and
\begin{eqnarray}\nonumber
\fl\mathfrak
C_1(a_k)\;&:=\;\frac{1}{a_k}\,+\,\sum_{n=1}^{\infty}\frac{2a_k}{a_k^2-4\pi^2n^2}\;=\;\frac{1}{a_k}
\,+\,\frac{\Psi\left(1-\frac{a_k}{2\pi}\right)-\Psi\left(1+\frac{a_k}{2\pi}\right)}{2\pi}
\;=\;\frac{1}{2}\cot\frac{a_k}{2}\\[5pt]
\fl\mathfrak
C_2(a_k)\;&:=\;\frac{1}{a_k^2}\,+\,\sum_{n=1}^{\infty}\frac{1}{(a_k+2\pi
n)^2}+\frac{1}{(a_k-2\pi
n)^2}\;=\;\frac{1}{a_k^2}\,+\,\frac{1}{4\sin^2\frac{a_k}{2}},
\end{eqnarray}

\noindent where $\Psi$ denotes the Digamma function
$\Psi(z)=\Gamma'(z)/\Gamma(z)$. Now we have to employ the
identities
\begin{eqnarray}\label{Gl:HadamardOfSin(z)}
\sin
z\;=\;z\prod_{n=1}^{\infty}\left(1-\frac{z^2}{\pi^2n^2}\right)
\end{eqnarray}

\noindent and
\begin{eqnarray}\label{Gl:HadamardOfSin}
\fl\quad\sin\left(\frac{z-a}{2}\right)\;=\;-\sin\left(\frac{a}{2}\right)\;\left(1-\frac{z}{a}\right)\;
\prod_{n=1}^{\infty}\left(1-\frac{z}{a+2\pi
n}\right)\left(1-\frac{z}{a-2\pi n}\right)
\end{eqnarray}

\noindent for $\frac{a}{2\pi}\notin \Z$. Equation
(\ref{Gl:HadamardOfSin(z)}) is well known and
(\ref{Gl:HadamardOfSin}) will be proved in the appendix.
Reinserting these two identities in (\ref{Gl:Gleichung1}), we
obtain:
\begin{eqnarray}
\fl \psi_{\d}(z)\;=\;\sin^m\frac{z}{2}\;\exp\left({mz^2\sum_{n=1}^{\infty}\frac{1}{4\pi^2n^2}}\right)\;\\[5pt]\nonumber
\;\times\;\rme^{C_0+C_1z+C_2z^2}\;\prod_{k}\frac{\sin\frac{z-a_k}{2}}{\sin\frac{-a_k}{2}}\;
\exp\left({\mathfrak C_1(a_k)z\,+\,\frac{1}{2}\mathfrak
C_2(a_k)z^2}\right).
\end{eqnarray}

\noindent The first exponential factor can be absorbed into a
redefinition of $C_2$:
\begin{eqnarray}\label{Gl:PeriodicityNicerForm}
\fl \psi_{\d}(z)\;=\;&\sin^m\frac{z}{2}\;\rme^{C_0+
C_1z+\widetilde{
C_2}z^2}\;\prod_{k}\frac{\sin\frac{z-a_k}{2}}{\sin\frac{-a_k}{2}}\;
\exp\left({\mathfrak C_1(a_k)z\,+\,\frac{1}{2}\mathfrak
C_2(a_k)z^2}\right).
\end{eqnarray}

\noindent Periodicity now demands that
\begin{eqnarray}\nonumber
\fl \psi_{\d}(z)\;=\;\rme^{2\pi \rmi \d}\psi(z+2\pi)\;=\;(-1)^m
\,\sin\frac{z}{2}\, \rme^{2\pi \rmi \d}\,\rme^{ C_0+ C_1(z+2\pi)+\widetilde{
C_2}(z+2\pi)^2}\;\\[5pt]
\fl\quad\qquad\quad\times\,\prod_{k}\left[-\frac{\sin\frac{z-a_k}{2}}{\sin\frac{-a_k}{2}}\;
\exp\left({\mathfrak C_1(a_k)(z+2\pi)\,+\,\frac{1}{2}\mathfrak
C_2(a_k)(z+2\pi)^2}\right)\right]\\[5pt]\nonumber
\fl\quad\qquad\;=\;\psi_{\d}(z)\,(-1)^m\,\rme^{2\pi
\rmi\d\,+\,2\pi C_1+4\pi^2\widetilde{C_2}+4\pi\widetilde{
 C_2}z}\,\prod_k\left[-\rme^{2\pi \mathfrak
C_1(a_k)+4\pi^2\mathfrak C_2(a_k)+4\pi \mathfrak C_2(a_k)z}\right]
\end{eqnarray}

\noindent so
\begin{eqnarray}\label{Gl:PeriodicityCondition}
 \fl 1\;=\;(-1)^m\,\rme^{2\pi \rmi\d\,+\,2\pi C_1+4\pi^2\widetilde{
C_2}+4\pi\widetilde{
C_2} z}\,\\[5pt]\nonumber
\times\;\prod_k\Bigg[-\exp\Big(2\pi \mathfrak
C_1(a_k)+2\pi^2\mathfrak C_2(a_k)+2\pi \mathfrak
C_2(a_k)z\Big)\Bigg].
\end{eqnarray}

\noindent Note that this equation holds for all $z\in\C$, also for
$z=0$. Hence, both
\begin{eqnarray}
\prod_k\Bigg[-\exp\Big({2\pi \mathfrak C_1(a_k)+2\pi^2\mathfrak
C_2(a_k)+2\pi \mathfrak C_2(a_k)z}\Big)\Bigg]
\end{eqnarray}

\noindent and

\begin{eqnarray}
\prod_k\Bigg[-\exp\Big({2\pi \mathfrak C_1(a_k)+2\pi^2\mathfrak
C_2(a_k)}\Big)\Bigg]
\end{eqnarray}

\noindent converge. Thus, also their quotient
\begin{eqnarray}
\prod_k\Bigg[\exp\Big({2\pi \mathfrak
C_2(a_k)z}\Big)\Bigg]\;=\;\exp\left({z\sum_k2\pi \mathfrak
C_2(a_k)}\right)
\end{eqnarray}

\noindent converges for each $z\in\C$, and can be pulled out of
the (possibly infinite) product in
(\ref{Gl:PeriodicityCondition}). So we get:
\begin{eqnarray}\nonumber
\fl 1\;=\;(-1)^m\,\rme^{2\pi \rmi\d\,+\,2\pi C_1+4\pi^2\tilde C_2}
\exp\left(4\pi \widetilde{ C_2}+\sum_k2\pi \mathfrak
C_2(a_k)\right)z\,\,\\[5pt]\label{Gl:PeriodicityCondition2}
\times\;\prod_k\Bigg[-\exp\Big({2\pi \mathfrak
C_1(a_k)+2\pi^2\mathfrak C_2(a_k)}\Big)\Bigg].
\end{eqnarray}

\noindent This is only possible, if
\begin{eqnarray}\label{Gl:PeriodicityFirstResult}
4\pi \widetilde{ C_2}+\sum_k2\pi \mathfrak C_2(a_k)\;=\;0,
\end{eqnarray}

\noindent as can be easily seen if one compares the growth
behavior on both sides of (\ref{Gl:PeriodicityCondition2}).
Inserting (\ref{Gl:PeriodicityFirstResult}) into
(\ref{Gl:PeriodicityCondition2}) yields
\begin{eqnarray}\label{Gl:ConditionForC_1}
\fl \rme^{-\rmi m\pi\,-\,2\pi \rmi\d\,-\,2\pi
C_1}\;=\;\prod_k\Bigg[-\exp\Big({2\pi\mathfrak
C_1(a_k)}\Big)\Bigg]\;=\;\prod_k\Bigg[-\exp\Big({\pi\cot\frac{a_k}{2}}\Big)\Bigg].
\end{eqnarray}

\noindent To proceed, we need a technical lemma:

\begin{Lemma}\label{Lem:EinzigesLemmaImArtikel}
Let $\{a_k\}$ be a sequence in $([0,2\pi)\times
\rmi\R)\backslash\{0\}$ of order 1. Let $\n_k\,:=\,{\rm sgn}\,{\rm
Im}\,a_k$ be the sign of the imaginary part of $a_k$ (with {\rm
sgn}\,0\;:=\;1). Then
\begin{eqnarray}\label{Gl:Lemma41}
\prod_k\Bigg[-\exp\Big({\pi\cot\frac{a_k}{2}}\Big)\Bigg]\;=\;\exp\left[\sum_k\pi\left(
\cot\frac{a_k}{2}\,+\,\n_k\rmi\right)\right]
\end{eqnarray}
\end{Lemma}

\noindent The proof of this lemma is rather technical and will be
delivered in \ref{app-proof}.
With the help of (\ref{Gl:Lemma41}),
(\ref{Gl:ConditionForC_1}) can be rewritten. If the exponentials
of two complex numbers are equal, the numbers themselves are equal
up to a multiple of $2\pi \rmi$. So, there is a number $l\in\Z$
such that:
\begin{eqnarray}
-\rmi m\pi\,-\,2\pi \rmi\d\,-\,2\pi
C_1\;=\;2\pi\sum_k\left(\n_k\frac{\rmi}{2}+\mathfrak
C_1(a_k)\right)\,+2\pi \rmi l.
\end{eqnarray}

\noindent Thus
\begin{eqnarray}\label{Gl:PeriodicitySecondResult}
\rme^{C_1z}\;=\;\rme^{-\rmi\frac{m}{2}z}\,\rme^{\rmi(l-\d
)z}\,\prod_k\exp\left[{-z\left(\n_k\frac{\rmi}{2}+\mathfrak
C_1(a_k)\right)}\right].
\end{eqnarray}

\noindent Inserting (\ref{Gl:PeriodicityFirstResult}) and
(\ref{Gl:PeriodicitySecondResult}) into
(\ref{Gl:PeriodicityNicerForm}) gives then:

\begin{eqnarray}\label{Gl:HusimiFinaleFormel}
\fl
\psi_{\d}(z)\;=\;\rme^{C+\rmi(l-\d)z}\;\left[\,\sin\frac{z}{2}\,\exp\left({-\rmi\frac{z}{2}}\right)\,\right]^m\,
\prod_{k}\;\left[\,\frac{\sin\frac{z-a_k}{2}}{\sin\frac{-a_k}{2}}\,\exp\left({-\n_k\rmi\frac{z}{2}}\right)\,\right],
\end{eqnarray}

\noindent which completes the proof.\\

  Note that the exponentials $\exp\left({-\n_k\rmi\frac{z}{2}}\right)$ in
formula (\ref{Gl:HusimiFinaleFormel}) can only be pulled out of
the product if the set of zeros $a_k$ is finite. If this is not
the case, the exponentials are needed for the infinite product to
converge, of which the particular choice of the sign $\n_k$
according to the imaginary part of $a_k$ takes care.

 This proposition shows that a state is (up to
normalization) completely determined by the zeros in its
Hadamard-decomposition and one further integer $l$. This is in
contrast to the case of quantum mechanics on the real line, where
instead of the integer $l$ one has to specify two complex numbers
$C_1$ and $C_2$.

From this one can immediately conclude that the only states whose
Husimi distributions are positive definite, are (up to
normalization) the basis vectors (\ref{Gl:Eigenvectors}). Let
$|\psi\rangle$ and $|\tilde{\psi}\rangle$ be two states such that
their Bargmann-Segal representations (\ref{Gl:Bargmann-Segal})
have the same zeros (with the same multiplicities), i.e.

\begin{eqnarray}
\psi_{\d}(z)=0\;\Leftrightarrow\;\tilde{\psi}_{\d}(z)=0.
\end{eqnarray}

\noindent By (\ref{Gl:HadamardPeriodic}) this is the case if and
only if $\tilde{\psi}_{\d}(z)=\exp(C+\rmi l z)\psi_{\d}(z)$ for
some $l\in\Z$. With the action of the ladder operators $\hat g$
and $\hat g^{\dag}$ (\ref{Gl:LadderOperators2}), one can
immediately conclude that, up to normalization,
\begin{eqnarray}
|\tilde\psi\rangle\;=\;\left(\hat g^{\dag}\right)^l\,|\psi\rangle.
\end{eqnarray}

\noindent So, since the Bargmann-Segal representation of the basis
vectors (\ref{Gl:Eigenvectors}) has no zeros and the basis vectors
are transformed into each other by application of $\hat g^{\dag}$,
as one can easily compute, every state whose Bargmann-Segal
representation has no zeros, is, up to normalization, a basis vector
$|n\rangle_{\d}$, $n\in\Z$. In particular, the Husimi distribution
of every complexifier coherent state $|z\rangle_{\d}$ has at least
one zero, in contrast to the situation for quantum mechanics on the
real line. Even more, since the Husimi-distribution of a coherent
state can, with the definition (\ref{Gl:CoherentStates}), be written
in terms of the third Jacobian theta function
(\ref{Gl:ThetaFunction}),

\begin{eqnarray}\label{Gl:CoherentStateAsThetaFunction}
\psi_{\d}^{z',t}(z)\;:=\;{}_{\d}\langle\,
z'\,|\,z\,\rangle_{\d}\;=\;\rme^{-\d^2s^2}\;\vartheta_3\left(\frac{2\rmi
s^2\d+\bar{z}'-z}{2\pi},\frac{\rmi s^2}{\pi}\right),
\end{eqnarray}

\noindent and the zeros of the theta function $\vartheta_3(z,\t)$
for fixed $\t$ are known \cite{APOSTOL} to be
\begin{eqnarray}
z_0=\left(k+\frac{1}{2}\right)+\left(m+\frac{1}{2}\right)\t\qquad\mbox{
for all }k,m\in\Z,
\end{eqnarray}
\noindent we immediately see that the Husimi-distribution of
the coherent states has even infinitely many zeros.\\

 As we have seen by (\ref{Gl:HadamardPeriodic}), one
needs considerably less effort to reconstruct the state by the
zeros of the Bargmann-Segal representation (and hence the
Husimi-distribution). Also, these zeros for states on the circle
behave quite differently than the ones for states on the real
line.

\section{Semiclassical Propagator}\label{Ch:Propagator}
\subsection{Semiclassical approximation}

\noindent One of the most widely acknowledged property of the
harmonic oscillator coherent states is the fact that they provide a
gateway to semiclassical analysis of quantum mechanical systems on
the real line. In particular, they can be used to approximate the
propagator between coherent states $K_{\R}(z_F,t_F,z_I,t_I)$ or
position eigenstates $K_{\R}(x_F,t_F,x_I,t_I)$ (see, e.g.
\cite{KECK,Pari03,Ribe04,Nova05,Ribe05,Pari06} and references
therein). In this section, we will show that the coherent states
presented in the last sections can be used to calculate the
propagator for quantum mechanical systems on the circle in a
semiclassical approximation in the same way. Although technically
more elaborate, there is not much conceptual difference between the
following derivation and the one in \cite{KECK}. This will
demonstrate that the $U(1)$-complexifier coherent states are in fact
useful for a semiclassical analysis. So, in what follows we will
compute the coherent state propagator
\begin{eqnarray}\label{Gl:CoherentPropagator}
K_{S^1}(z_F,\,\t,\,z_I,\,0)\;=\;{}_{\d}\langle
z_{F}|\rme^{-\frac{\rmi}{\hbar}\hat H\t}|z_I\rangle_{\d},
\end{eqnarray}

\noindent since the propagators in the "angle"-, momentum- or any
mixed representation can be derived from it.

Although we are working with the semiclassical approximation of the
propagator, an exact expression can be written down for the
particular case of
the freely moving particle on the circle (see \cite{KASTRUP}).\\

\noindent We start with expanding the exponential. For large $N$
we have
\begin{eqnarray}\label{Gl:PropagatorExponentialExpansion}
\;{}_{\d}\langle z_{F}|\rme^{-\frac{\rmi}{\hbar}\hat
H\t}|z_I\rangle_{\d}\;\approx\; {}_{\d}\langle
z_{F}|\left(\mathbbm{1}\;-\;\rmi\frac{\t}{N\hbar}\hat
H\right)^N|z_I\rangle_{\d}.
\end{eqnarray}

\noindent Using the completeness of the coherent states
(\ref{Gl:CoherernOvercomplete}), we get

\begin{eqnarray}\nonumber
\fl{}_{\d}\langle z_{F}|\rme^{-\frac{\rmi}{\hbar}\hat
H\t}|z_I\rangle_{\d}\;\approx\;&\frac{1}{\sqrt{\pi
s^2\hbar^2}}\;\int_{\R^{N+1}}\rmd p_{N}\ldots
\rmd p_0\,\int_{[0,2\pi]^{N+1}}\frac{\rmd\phi_{N}}{2\pi}\cdots\frac{\rmd\phi_{0}}{2\pi}\\[5pt]\label{Gl:PropagatorBasicFormula}
\fl&\times\;\prod_{k=0}^{N}\left[\exp{\left(\frac{z_k-\bar
z_k}{2s}\right)^2}\right]\;{}_{\d}\langle
z_{F}|z_N\rangle_{\d}\;{}_{\d}\langle
z_0|z_I\rangle_{\d}\\[5pt]\nonumber
\fl&\times\prod_{k=0}^{N-1}\left[{}_{\d}\langle
z_{k+1}|z_k\rangle_{d}\;\left(\mathbbm{1}\,-\,\frac{\rmi\t}{N\hbar}\H(\bar
z_{k+1},\,z_k)\right)\right],
\end{eqnarray}

\noindent where the function $\H$ defined by:
\begin{eqnarray}\label{Gl:PropagatorHamiltonian}
\H(w,z)\;:=\;\frac{{}_{\d}\langle\bar w|\hat
H|z\rangle_{\d}}{{}_{\d}\langle \bar w|z\rangle_{\d}}
\end{eqnarray}

\noindent  is holomorphic in both variables.\\

\noindent With (\ref{Gl:CoherentOverlap}) we get

\begin{eqnarray}\nonumber
\fl{}_{\d}\langle z_{F}|\rme^{-\frac{\rmi}{\hbar}\hat
H\t}|z_I\rangle_{\d}\;\approx\;
\frac{1}{(s^2\hbar)^{N+1}}\sqrt\frac{\pi}{s^2}
\;\int_{\R^{N+1}}\rmd p_{N}\ldots \rmd
p_0\,\int_{[0,2\pi]^{N+1}}\frac{\rmd
\phi_{N}}{2\pi}\cdots\frac{\rmd
\phi_{0}}{2\pi}\\[5pt]\label{Gl:PropagatorBasicFormula2}
\fl\times\;\prod_{k=0}^{N}\left[\exp\left(\frac{z_k-\bar
z_k}{2s}\right)^2\right]\;\sum_{n\in\Z} \exp\left[2\pi \rmi n
\d\;-\;\left( \frac{\bar
z_{F}-z_{N}-2\pi n}{2s}\right)^2\right]\\[5pt]\nonumber
\fl\times \sum_{n\in\Z} \exp\left[2\pi \rmi n \d\;-\;\left(\frac{\bar
z_{0}-z_I-2\pi n}{2s}\right)^2\right]\\[5pt]\nonumber
\fl\times\prod_{k=0}^{N-1}\sum_{n\in\Z} \exp\left[2\pi \rmi n
\d\;-\;\left(\pi n - \frac{\bar z_{k+1}-z_k-2\pi
n}{2s}\right)^2\;-\;\frac{\rmi\t}{N\hbar}\H(\bar
z_{k+1},\,z_k)\right],
\end{eqnarray}

\noindent where
\begin{eqnarray}
\left(\mathbbm{1}\,-\,\frac{\rmi\t}{N\hbar}\H(\bar
z_{k+1},\,z_k)\right)\approx\;\exp\left(-\frac{\rmi\t}{N\hbar}\H(\bar
z_{k+1},\,z_k)\right)
\end{eqnarray}

\noindent has been used. With
\begin{eqnarray}
\prod_{k=0}^{N-1}\left(\sum_{n\in\Z}\;f(n)\right)\;=\;\sum_{n_1\in\Z}\cdots\sum_{n_{N}\in\Z}\prod_{k=0}^{N-1}f(n_{k+1}),
\end{eqnarray}

\noindent formula (\ref{Gl:PropagatorBasicFormula2}) can be
rewritten as follows:
\begin{eqnarray}\nonumber
\fl{}_{\d}\langle z_{F}|\rme^{-\frac{\rmi}{\hbar}\hat
H\t}|z_I\rangle_{\d}\;\approx\;&\frac{1}{(s^2\hbar)^{N+1}}\sqrt\frac{\pi}{s^2}
\;\int_{\R^{N+1}}\rmd p_{N}\ldots
\rmd p_0\,\int_{[0,2\pi]^{N+1}}\frac{\rmd \phi_{N}}{2\pi}
\cdots\frac{\rmd \phi_{0}}{2\pi}\\[5pt]\label{Gl:PropagatorBasicFormula3}
\fl&\times\;\sum_{n_0\in\Z}\sum_{n_1\in\Z}\cdots\sum_{n_{N+1}\in\Z}\exp\Big[f^{(\vec
n)}(\vec{\bar{z}},\,\vec z)\Big],
\end{eqnarray}

\noindent where
\begin{eqnarray}
f^{(\vec n)}(\vec{\bar{z}},\,\vec
z)\;=\;f^{(n_0,\ldots,n_{N+1})}(\bar z_0,\ldots,\bar
z_N,z_0,\ldots,z_N)
\end{eqnarray}

\noindent is given by
\begin{eqnarray}\nonumber
\fl f^{(n_0,\ldots,n_{N+1})}(\bar z_0,&\ldots,\bar
z_N,z_0,\ldots,z_N)\;=\;2\pi \rmi\d
\left(\sum_{k=0}^{N+1}n_k\right)\;+\;\sum_{k=0}^{N}\left(\frac{z_k-\bar
z_k}{2s}\right)^2\\[5pt]\label{Gl:PropagatorFormOfF}
\fl &\quad-\;\left(\frac{\bar z_0-z_I-2\pi
n_0}{2s}\right)^2\;-\;\sum_{k=0}^{N-1}\left(\frac{\bar z_{k+1}-
z_k-2\pi
n_{k+1}}{2s}\right)^2\\[5pt]\nonumber
\fl &\quad-\;\left(\frac{\bar z_F-z_N-2\pi
n_{N+1}}{2s}\right)^2\;-\frac{\rmi\t}{N\hbar}\sum_{k=0}^{N-1}\H(\bar
z_{k+1},\,z_k).
\end{eqnarray}

\noindent This formula can be simplified considerably by using the
following property of $f^{(\vec n)}(\vec{\bar{z}},\,\vec z)$:
\begin{eqnarray}\label{Gl:PropagatorPropertyOfF}
\fl f^{(n_0,\ldots,n_{N+1})}(\bar z_0,\ldots,\bar
z_k+2\pi,\ldots,\bar z_N,z_0,\ldots,
z_k+2\pi,\ldots,z_N)\\[5pt]\nonumber
\;=\;f^{(n_0,\ldots,n_{k}-1,n_{k+1}+1,\ldots,n_{N+1})}(\bar
z_0,\ldots,\bar z_N,z_0,\ldots,z_N),
\end{eqnarray}

\noindent which can be readily seen from the explicit form
(\ref{Gl:PropagatorFormOfF}) of $f^{(\vec n)}(\vec{\bar{z}},\,\vec
z)$ and the fact that $\H$ is periodic in both variables
(\ref{Gl:PropagatorHamiltonian}), since the coherent states are.\\

\noindent We then have
\begin{eqnarray}\label{Gl:PropagatorTranslationTrick}
\fl\sum_{n_0\in\Z}\sum_{n_1\in\Z}\int\limits_{\,[0,2\pi]^2}\frac{\rmd\phi_1}{2\pi}\frac{\rmd\phi_0}{2\pi}
\exp\Big[f^{(n_0,\ldots,n_{N+1})}(\bar z_0,\ldots,\bar
z_n,z_0,\ldots,z_N)\Big]\nonumber\\[5pt]\nonumber
\fl\;\!=\;\!\!\!\sum_{n_0\in\Z}\sum_{n_1\in\Z}\int\limits_{[0,2\pi]^2}\!\!\frac{\rmd\phi_1}{2\pi}\frac{\rmd\phi_0}{2\pi}
\exp\Big[f^{(0,n_1\!-\!n_0,\ldots,n_{N+1})}(\bar z_0\!+\!2\pi
n_0,\ldots,\bar
z_n,z_0\!+\!2\pi n_0,\ldots,z_N)\Big]\\[5pt]\nonumber
\fl\;\!=\;\!\!\!\sum_{n_0\in\Z}\sum_{n_1\in\Z}\int\limits_{\,[0,2\pi]^2}\frac{\rmd\phi_1}{2\pi}\frac{\rmd\phi_0}{2\pi}
\exp\Big[f^{(0,n_1,\ldots,n_{N+1})}(\bar z_0+2\pi n_0,\ldots,\bar
z_n,z_0+2\pi n_0,\ldots,z_N)\Big]\\[5pt]
\fl\;\!=\;\!\!\!\sum_{n_1\in\Z}\int\limits_{\,[0,2\pi]}\frac{\rmd\phi_1}{2\pi}\int_{\R}\frac{\rmd\phi_0}{2\pi}
\exp\Big[f^{(0,n_1\ldots,n_{N+1})}(\bar z_0,\ldots,\bar
z_n,\ldots,z_N)\Big].
\end{eqnarray}

\noindent In the second step we have used the invariance under
shifting the summation index $n_1$. The trick used in
(\ref{Gl:PropagatorTranslationTrick}) can now be carried on over
all the $\phi_k$ to transform the integrations over $[0,2\pi]$
into integrations over $\R$ while getting rid of the summations.
The last summation over $n_{N+1}$, however, can not be eliminated
this way, and one obtains:
\begin{eqnarray}\nonumber
\fl {}_{\d}\langle z_{F}|\rme^{-\frac{\rmi}{\hbar}\hat
H\t}|z_I\rangle_{\d}\;\approx\;
\frac{1}{(s^2\hbar)^{N+1}}\sqrt\frac{\pi}{s^2}\;
\sum_{n\in\Z}\int_{\R^{N+1}}\rmd p_{N}\ldots
\rmd p_0\,\int_{\R^{N+1}}\frac{\rmd\phi_{N}}{2\pi}\cdots\frac{\rmd\phi_{0}}{2\pi}\\[5pt]\label{Gl:PropagatorBasicFormula4}
\times\;\exp\Big[f^{(0,0,\ldots,0,n)}(\vec{\bar{z}},\,\vec
z)\Big].
\end{eqnarray}

\noindent The integrand now takes a much simpler form:
\begin{eqnarray}\label{Gl:PropagatorFormOfFSimplified}
\fl f^{(0,\,0,\,\ldots,n)}&(\vec{\bar{z}},\,\vec z) \;=\;2\pi \rmi
n\d\;+\; \sum_{k=0}^{N}\left(\frac{z_k-\bar
z_k}{2s}\right)^2-\;\left(\frac{\bar
z_0-z_I}{2s}\right)^2\\[5pt]\nonumber
\fl &\;-\;\sum_{k=0}^{N-1}\left(\frac{\bar z_{k+1}-
z_k}{2s}\right)^2\;-\;\left(\frac{\bar z_F-z_N-2\pi
n}{2s}\right)^2\;-\frac{\rmi\t}{N\hbar}\sum_{k=0}^{N-1}\H(\bar
z_{k+1},\,z_k).
\end{eqnarray}

\noindent We are now able to perform the semiclassical
approximation. It assumes that the main part of the integral
(\ref{Gl:PropagatorBasicFormula4}) comes from the stationary
points of the integrand. These points are characterized by the
condition that the first derivative of
(\ref{Gl:PropagatorFormOfFSimplified}) vanishes:
\begin{eqnarray}\nonumber
0=\frac{\del f^{(\vec 0,n)}}{\del z_k}&=\frac{\bar
z_{k+1}-\bar
z_k}{2s^2}\,-\,\frac{\t}{N}\frac{\rmi}{\hbar}\del_2\H(\bar
z_{k+1},\,z_k)\;,\quad k=0,\ldots,N-1\\[5pt]\label{Gl:PropagatorFirstDerivatives}
0=\frac{\del f^{(\vec 0,n)}}{\del z_N}&=\frac{2\pi n -
\bar
z_F+\bar z_N}{2s^2}\\[5pt]\nonumber
0=\frac{\del f^{(\vec 0,n)}}{\del \bar z_{k+1}}&=\frac{
z_{k}-
z_{k+1}}{2s^2}\,-\,\frac{\t}{N}\frac{\rmi}{\hbar}\del_1\H(\bar
z_{k+1},\,z_k)\;,\quad k=0,\ldots,N-1\\[5pt]\nonumber
0=\frac{\del f^{(\vec 0,n)}}{\del \bar
z_0}&=\frac{z_I-z_0}{2s^2},
\end{eqnarray}

\noindent where we have used the notation
\begin{eqnarray}
\del_1\H(w,z)\;=\;\frac{\del\H}{\del
w}(w,z),\qquad\del_2\H(w,z)\;=\;\frac{\del\H}{\del z}(w,z).
\end{eqnarray}

\noindent Let the points where
(\ref{Gl:PropagatorFirstDerivatives}) are satisfied be called
$(\vec {\bar w},\,\vec w)$.\\

Note that there may be more than one set of complex numbers
satisfying (\ref{Gl:PropagatorFirstDerivatives}) for every $N$ and
$n$. This resembles the fact that there may be more than one
classical trajectory from one point on the circle to another with
fixed winding number. We will keep this in mind, but refrain from
introducing a special notation to keep track of the different
paths, in order not to overburden the formulae.\\

\noindent We now expand $f^{(\vec 0,n)}$ up to second order:
\begin{eqnarray}\label{Gl:PropagatorExpansionOfF}
f^{(\vec 0,n)}(\vec{\bar{z}},\,\vec{z})\;\approx\;&f^{(\vec
0,n)}(\vec{\bar{w}},\,\vec{w})\;+\;\frac{1}{2}\frac{\del^2f^{(\vec
0,n)}}{\del z_{k}\del
z_{l}}(\vec{\bar{w}},\,\vec{w})z^kz^l\\[5pt]\nonumber
&\;+\;\frac{\del^2f^{(\vec 0,n)}}{\del z_{k}\del \bar
z_{l}}(\vec{\bar{w}},\,\vec{w})z^k\bar
z^l\;+\;\frac{1}{2}\frac{\del^2f^{(\vec 0,n)}}{\del \bar z_{k}\del
\bar z_{l}}(\vec{\bar{w}},\,\vec{w})\bar z^k\bar z^l.
\end{eqnarray}

\noindent Inserting (\ref{Gl:PropagatorExpansionOfF}) into
(\ref{Gl:PropagatorBasicFormula4}) enables us to evaluate the
integral in (\ref{Gl:PropagatorBasicFormula4}) explicitly, for we
will have to deal with ordinary Gaussian integrals only. From
(\ref{Gl:PropagatorFirstDerivatives}) we can compute the second
derivatives of $f^{(\vec 0,n)}$:
\begin{eqnarray}\nonumber
\fl \frac{\del^2f^{(\vec 0,n)}}{\del z_k\del
z_l}&=-\frac{\t}{N}\frac{\rmi}{\hbar}\del_2^2\H(\bar
z_{k+1},\,z_k)\,\d_{kl}\;\qquad k,l=0,\ldots,N-1,\\[5pt]\nonumber
\fl \frac{\del^2f^{(\vec 0,n)}}{\del z_N\del z_k}&=0\;\qquad
k=0,\ldots,N,\\[5pt]\nonumber
\fl \frac{\del^2f^{(\vec 0,n)}}{\del z_k\del \bar
z_{l+1}}&=-\frac{1}{2s^2}\d_{k,l+1}\!+\!\left(\frac{1}{2s^2}\!-\!\frac{\t}{N}\frac{\rmi}{\hbar}\del_1\del_2\H(\bar
z_{k+1},\,z_k)\right)\d_{kl}\,,\quad k,l=0,\ldots,N-1,\\[5pt]\nonumber
\fl \frac{\del^2f^{(\vec 0,n)}}{\del z_k\del \bar
z_{0}}&=-\frac{1}{2s^2}\,\d_{0k}\;\qquad
k=0,\ldots,N,\\[5pt]\label{Gl:PropagatorSecondDerivatives}
\fl \frac{\del^2f^{(\vec 0,n)}}{\del z_N\del \bar
z_{k}}&=-\frac{1}{2s^2}\,\d_{0N}\;\qquad
k=0,\ldots,N,\\[5pt]\nonumber
\fl \frac{\del^2f^{(\vec 0,n)}}{\del \bar z_{k+1}\del \bar
z_{l+1}}&=-\frac{\t}{N}\frac{\rmi}{\hbar}\del_1^2\H(\bar
z_{k+1},\,z_k)\,\d_{kl}\;\qquad k,l=0,\ldots,N-1,\\[5pt]\nonumber
\fl \frac{\del^2f^{(\vec 0,n)}}{\del \bar z_0\del \bar
z_{k}}&=0\;\qquad k=0,\ldots,N.
\end{eqnarray}

\noindent Inserting (\ref{Gl:PropagatorExpansionOfF}) and
(\ref{Gl:PropagatorSecondDerivatives}) into
(\ref{Gl:PropagatorBasicFormula4}), we obtain:
\begin{eqnarray}\nonumber
\fl{}_{\d}\langle z_{F}|\rme^{-\frac{\rmi}{\hbar}\hat
H\t}|z_I\rangle_{\d}\;\approx\;
\sqrt\frac{\pi}{s^2}\frac{1}{(\hbar
s^2)^{N+1}}\;\sum_{n\in\Z}\rme^{f^{(\vec
0,n)}(\vec{\bar{w}},\,\vec{w})}\!\!\!\int\limits_{\R^{N+1}}\rmd p_{N}\ldots
\rmd p_0\!\!\!\int\limits_{\R^{N+1}}\frac{\rmd\phi_{N}}{2\pi}\cdots\frac{\rmd\phi_{0}}{2\pi}\\[5pt]\nonumber
\times\exp\Bigg[-\frac{1}{2}\sum_{k=0}^{N-1}\frac{\t}{N}\frac{\rmi}{\hbar}\del^2_2\H(\bar
w_{k+1},\,w_k)\,z_k^2\;-\;\frac{1}{2s^2}\sum_{k=0}^Nz_k\bar
z_k\\[5pt]\nonumber
+\;\sum_{k=0}^{N-1}\left(\frac{1}{2s^2}-\frac{\t}{N}\frac{\rmi}{\hbar}\del_1\del_2\H(\bar
w_{k+1},\,w_k)\right)z_k\bar z_{k+1} \\[5pt]\label{Gl:PropagatorBasicFormula5}
-\;\frac{1}{2}\sum_{k=0}^{N-1}\frac{\t}{N}\frac{\rmi}{\hbar}\del^2_1\H(\bar
w_{k+1},\,w_k) \bar z_{k+1}^2\Bigg].
\end{eqnarray}

\noindent For $z=\phi+{i}p/{\hbar}$, the Gaussian integral
\begin{eqnarray}\nonumber
\fl \frac{1}{2\pi\hbar}\int_{\R^2}\rmd\phi\,\rmd
p\,\exp\Big(a_1z^2+a_2\bar
z^2+a_3\bar zz+b_1z+b_2\bar z\Big)\\[5pt]\label{Gl:PropagatorGaussianIntegral}
=\;\frac{1}{2}\;\frac{1}{\sqrt{a_3^2-4a_1a_2}}\,\exp\left(\frac{a_1b_2^2+a_2b_1^2-a_3b_1b_2}{a_3^2-4a_1a_2}\right)
\end{eqnarray}

\noindent is used to integrate successively over the
$\rmd\phi_k\rmd p_k$, starting with $k=0$. From
(\ref{Gl:PropagatorBasicFormula5}), we can read off the parameters
for this integration:
\begin{eqnarray}\label{Gl:PropagatorFirstSetParameters}
\fl
a_1\;=\;-\frac{1}{2}\frac{\t}{N}\frac{\rmi}{\hbar}\del_2^2\H(\bar
w_1,\,w_0),\qquad a_2\;=\;0\;=:\;X_0,\qquad a_3\;=\;-\frac{1}{2s^2},\\[5pt]\nonumber
b_1\;=\;\left(\frac{1}{2s^2}-\frac{\t}{N}\frac{\rmi}{\hbar}\del_1\del_2\H(\bar
w_{1},\,w_0)\right)\bar z_1,\qquad b_2\;=\;0.
\end{eqnarray}

\noindent The integration yields (keeping $X_0=0$):

\begin{eqnarray}\nonumber
\fl{}_{\d}\langle z_{F}|\rme^{-\frac{\rmi}{\hbar}\hat
H\t}|z_I\rangle_{\d}\;\approx\;
\sqrt\frac{\pi}{s^2}\frac{1}{(\hbar s^2)^{N+1}} \rme^{f^{(\vec
0,n)}(\vec{\bar{w}},\,\vec{w})}\;\int_{\R^{N}}\rmd p_{N}\ldots
\rmd p_1\,\int_{\R^{N}}\frac{\rmd\phi_{N}}{2\pi}\cdots\frac{\rmd\phi_{1}}{2\pi}\\[5pt]\nonumber
\fl\qquad \times\frac{1}{\sqrt{1+8s^4\frac{\t}{N}\frac{\rmi}{\hbar}\del_2^2H(\bar
w_1,\,w_0)\,X_0}}\;\exp\left[\frac{\left(1-2s^2\frac{\t}{N}\frac{\rmi}{\hbar}\del_1\del_2\H(\bar
w_1,\,w_0)\right)^2X_0}{1+8s^4\frac{\t}{N}\frac{\rmi}{\hbar}\del_2^2H(\bar
w_1,\,w_0)\,X_0}\right]
\\[5pt]\nonumber
\fl\qquad \times\exp\Bigg[-\frac{1}{4s^2}z_N^2-\frac{1}{2}\sum_{k=1}^{N-1}\frac{\t}{N}\frac{\rmi}{\hbar}\del^2_2\H(\bar
w_{k+1},\,w_k)\,z_k^2\;-\;\frac{1}{2s^2}\sum_{k=1}^Nz_k\bar
z_k\\[5pt]\label{Gl:PropagatorBasicFormulaAfterFirstIntegration}
\fl \qquad\qquad  +\;\sum_{k=1}^{N-1}\left(\frac{1}{2s^2}-\frac{\t}{N}\frac{\rmi}{\hbar}\del_1\del_2\H(\bar
w_{k+1},\,w_k)z_k\bar z_{k+1} \right)\\[5pt]\nonumber
\fl\qquad\qquad  -\;\frac{1}{2}\sum_{k=0}^{N-1}\frac{\t}{N}\frac{\rmi}{\hbar}\del^2_1\H(\bar
w_{k+1},\,w_k) \bar z_{k+1}^2\Bigg].
\end{eqnarray}

\noindent Now one can read off the next set of parameters
$a_1,a_2,a_3,b_1$ and $b_2$ to perform the subsequent integration
over $\rmd\phi_1\rmd p_1$ according to
(\ref{Gl:PropagatorGaussianIntegral}). The final result is
\begin{eqnarray}\label{Gl:PropagatorBasicFormulaAfterLastIntegration}
\fl {}_{\d}\langle z_{F}|\rme^{-\frac{\rmi}{\hbar}\hat
H\t}|z_I\rangle_{\d}\approx&
\sqrt\frac{\pi}{s^2}\;\sum_{n\in\Z}\rme^{f^{(\vec
0,n)}(\vec{\bar{w}},\,\vec{w})} \prod_{k=0}^{N-1}
\frac{1}{\sqrt{1+8s^4\frac{\t}{N}\frac{\rmi}{\hbar}\del_2^2\H(\bar
w_{k+1},\,w_k)\,X_k}},
\end{eqnarray}

\noindent where the $X_k$ are determined by the following
recursion relation:
\begin{eqnarray}\label{Gl:PropagatorRecursionRelation}
\fl X_0\;=\;0\\[5pt]\nonumber
\fl
X_{k+1}\;=\;\left(\frac{\left(1-2s^2\frac{\t}{N}\frac{\rmi}{\hbar}\del_1\del_2\H(\bar
w_{k+1},\,w_k)\right)^2X_k}{1+8s^4\frac{\t}{N}\frac{\rmi}{\hbar}\del_2^2H(\bar
w_{k+1},\,w_k)\,X_k}\right)\;-\;\frac{1}{2}\frac{\t}{N}\frac{\rmi}{\hbar}\del_1^2\H(\bar
w_{k+1},\,w_k).
\end{eqnarray}

\subsection{The continuum limit}

\noindent We now perform the limit $N\to\infty$ in
(\ref{Gl:PropagatorBasicFormulaAfterLastIntegration}). The
sequences $\bar w_{k},\,w_k$ become functions $\bar w(t),\,w(t)$
with $t$ ranging from $0$ to $\t$ and conditions
(\ref{Gl:PropagatorFirstDerivatives}) turn into the Hamiltonian
equations:
\begin{eqnarray}\label{Gl:PropagatorHamiltonEquations}
&\dot w\;=\;-2s^2\frac{\rmi}{\hbar}\del_1\H(\bar w,\,w),\qquad &w(0)\;=\;z_I\\[5pt]\nonumber
&\dot{\bar w}\;=\;2s^2\frac{\rmi}{\hbar}\del_2\H(\bar
w,\,w),\qquad &\bar w(\t)\;=\;\bar z_F-2\pi n.
\end{eqnarray}

\noindent At this point, we encounter the same phenomenon that
occurs in the case of the coherent state propagator of ordinary
quantum mechanics. Equations
(\ref{Gl:PropagatorHamiltonEquations}) are the Hamiltonian
equations for the system with the Hamiltonian $\H$, and the
boundary conditions fix the starting point $z_I$ and the endpoint
$z_F-2\pi n$ in phase space. But the trajectory is already fixed
by initial condition, and a solution $w(t)$ starting at $w(0)=z_I$
will most likely never go through $w(\t)=z_F-2\pi n$. The solution
to this problem is as follows: One has to give up the condition
that $w(t)$ and $\bar w(t)$ are to be complex conjugate to each
other. Defining new variables
\begin{eqnarray}
u\;:=\;w,\qquad v\;:=\;\bar w
\end{eqnarray}

\noindent one can see that the integral
(\ref{Gl:PropagatorBasicFormula4}) is complex analytic in $u$ and
$v$. Thus one can shift the integration plane defined by
$u\,=\,\bar v$ to another plane, where
(\ref{Gl:PropagatorHamiltonEquations}) actually has solutions, but
$u=\bar v$ is no longer guaranteed. Details can be found in
\cite{KECK}. This means that we have to solve the following set of
differential equations
\begin{eqnarray}\nonumber
\dot u\;&=\;-2s^2\frac{\rmi}{\hbar}\del_1\H(v,\,u)
\ ,\quad
\dot{v}\;&=\;2s^2\frac{\rmi}{\hbar}\del_2\H(v,\,u)\label{Gl:PropagatorHamiltonEquationsInUandV}
\end{eqnarray}
with boundary conditions
\begin{eqnarray}\label{Gl:PropagatorHamiltonEquationsBoundaryInUandV}
u(0)&\;=\;z_I
\ ,\quad
v(\t)&\;=\;\bar z_F-2\pi n.
\end{eqnarray}

\noindent Note that now the values of $u(\t)$ and $v(0)$ are not
directly given by $v(\t)$ and $u(0)$, in particular we do
\emph{not}, in general, have $v(0)= \overline{u(0})$ and
$u(\t)=\overline{ v(\t)}$. Rather, $v(0)$ and $u(\t)$ have to be
computed by solving (\ref{Gl:PropagatorHamiltonEquationsInUandV}),
they can hence be understood as functions of the in initial
conditions $u(0)=z_I$
and $v(\t)=\bar z_F-2\pi n$.\\

We now investigate the continuum limit for the different
factors in (\ref{Gl:PropagatorBasicFormulaAfterLastIntegration}),
starting with the exponential:
\begin{eqnarray}\nonumber
\fl{f^{(\vec 0,n)}(\vec{v},\,\vec{u})}&=2\pi n\d
\rmi+\sum_{k=0}^{N}\left(\frac{u_k-v_k}{2s}\right)^2-\sum_{k=0}^{N-1}\left[\left(\frac{u_k-v_{k+1}}{2s}\right)^2
+\frac{\rmi}{\hbar}\frac{\t}{N}\H(v_{k+1},\,u_k)\right]\\[5pt]\label{Gl:PropagatorLimitSFactor}
\fl&\quad-\;\left(\frac{v_0-z_I}{2s}\right)^2\;-\left(\frac{\bar z_F-2\pi n-u_N}{2s}\right)^2\\[5pt]\nonumber
\fl&=\;2\pi  n \d
\rmi\;+\;\sum_{k=0}^{N-1}\left[\frac{v_{k+1}-v_k}{4s^2}u_k-v_{k+1}\frac{u_{k+1}-u_k}{4s^2}
\,-\,\frac{\rmi\,\t}{N\hbar}\H(v_{k+1},\,u_k)\right]\\[5pt]\nonumber
\fl&\quad-\left(\frac{v_0-z_I}{2s}\right)^2-\left(\frac{\bar
z_F-2\pi n-u_N}{2s}\right)^2+
\frac{v_0(v_0-u_0)}{4s^2}+\frac{u_N(u_N-v_N)}{4s^2}\\[5pt]\label{Gl:PropagatorLimitSFactorLimit}
\fl(N\to\infty)\quad &=\;2\pi n\d
\rmi\;+\;\int_0^{\t}dt\,\left[\frac{\dot{v}u-\dot u
v}{4s^2}-\frac{\rmi}{\hbar}\H(v,\,u)\right]\;-\;\left(\frac{v(0)-z_I}{2s}\right)^2\\[5pt]\nonumber
\fl&\quad-\left(\frac{\bar z_F-2\pi
n-u(\t)}{2s}\right)^2+\frac{v(0)\big(v(0)-z_I\big)}{4s^2}+\frac{u(\t)\big(u(\t)-\bar
z_F+2\pi n\big)}{4s^2}.
\end{eqnarray}

\noindent We turn to the  difference equation
(\ref{Gl:PropagatorRecursionRelation}). As $N\to\infty$,
(\ref{Gl:PropagatorRecursionRelation}) becomes a differential
equation. We use the expansion
${(1+bx)^{-1}(1-ax)^2}=1-(2a+b)x+O(x^2)$. As $N$ becomes large,
one eventually gets, up to $O(N^{-2})$,
\begin{eqnarray}
\fl
X_{k+1}\;=\;X_k\;-\;X_k\Bigg(4s^2\del_1\del_2\H(v_{k+1},\,u_k)+8s^4\del_2^2H(v_{k+1},\,u_k)\,X_k\Bigg)
\frac{\t}{N}\frac{\rmi}{\hbar}\\[5pt]\nonumber
\;-\;\frac{1}{2}\frac{\t}{N}\frac{\rmi}{\hbar}\del_1^2\H(v_{k+1},\,u_k).
\end{eqnarray}

\noindent As $N\to\infty$, this becomes
\begin{eqnarray}\label{Gl:PropagatorDGLForX}
\dot
X\;=\;-4s^2X\frac{\rmi}{\hbar}\del_1\del_2H\,-\,8s^4X^2\frac{\rmi}{\hbar}\del_2^2H\,-\,\frac{\rmi}{2\hbar}\del_1^2H,
\end{eqnarray}

\noindent with boundary condition
\begin{eqnarray}\label{Gl:PropagatorInitialConditionForX}
X(0)\;=\;0.
\end{eqnarray}

\noindent Furthermore, we get
\begin{eqnarray}\nonumber
\fl\prod_{k=0}^{N-1}&
\frac{1}{\sqrt{1+8s^4\frac{\t}{N}\frac{\rmi}{\hbar}\del_2^2\H(v_{k+1},\,u_k)X_k}}\\[5pt]\label{Gl:PropagatorLimitXFactor}
\fl&\;=\;\exp\left[-\frac{1}{2}\sum_{k=0}^{N-1}\ln\Big(1+8s^4\frac{\t}{N}\frac{\rmi}{\hbar}
\del_2^2\H(v_{k+1},\,u_k)X_k\Big)\right]\\[5pt]\nonumber
\fl&\;=\;\exp\left[\sum_{k=0}^{N-1}\left(-4\frac{\rmi}{\hbar}s^4\del_2^2\H(v_{k+1},\,u_k)X_k\;+\;O(N^{-2})\right)
\right]\\[5pt]\nonumber
\fl(N\to\infty)
&\;=\;\exp\left(-4s^4\frac{\rmi}{\hbar}\int_0^{\t}dt\,\del_2^2\H\big(v(t),\,u(t)\big)\,X(t)\right).
\end{eqnarray}

\noindent To solve the differential equations
 (\ref{Gl:PropagatorDGLForX}), we
 perturb the boundary data
(\ref{Gl:PropagatorHamiltonEquationsBoundaryInUandV}) around a
given solution $v(t),\, u(t)$ of
(\ref{Gl:PropagatorHamiltonEquationsInUandV}) via
\begin{eqnarray}\label{Gl:PropagatorInitialPerturbations}
\tilde u(0)\,=\,u(0)\,+\,\d u(0),\qquad
\tilde{v}(\t)\,=\,v(\t)\,+\,\d v(\t),
\end{eqnarray}

\noindent where one has to determine the difference of this
solution with the original, i.e.~the evolution of the
perturbations $\d u(t),\,\d v(t)$. In first order of the
perturbation this yields
\begin{eqnarray}\label{Gl:PropagatorEvolutionOfPerturbations}
\d\dot u\;&=\;-2\rmi s^2\del_1^2\H(v,\,u)\,\d v\,-\,2\rmi s^2\del_1\del_2\H(v,\,u)\,\d u\\[5pt]\nonumber
\d \dot{v}\;&=\;2\rmi s^2\del_1^2\H(v,\,u)\,\d u\,+\,2\rmi
s^2\del_1\del_2\H(v,\,u)\,\d v.
\end{eqnarray}

\noindent Defining now
\begin{eqnarray}
X(t)\;:=\;\frac{1}{4s^2}\frac{\d u(t)}{\d v(t)},
\end{eqnarray}

\noindent one the finds with
(\ref{Gl:PropagatorEvolutionOfPerturbations}):
\begin{eqnarray}
\dot X\;&=\;\frac{1}{4s^2}\frac{\d \dot u}{\d
v}\,-\,\frac{1}{4}\frac{\d u}{(\d v)^2}\,\d \dot{v}\\[5pt]\nonumber
&=\;-\frac{\rmi}{2\hbar}\,\del_1^2\H(v,\,u)\,-\,\frac{\rmi}{\hbar}\frac{\d
u}{\d
v}\,\del_1\del_2\H(v,\,u)\,-\,\frac{\rmi}{2\hbar}\left(\frac{\d
u}{\d v}\right)^2\!\del_2^2\H(v,\,u)\\[5pt]\nonumber
&=\;-\frac{\rmi}{2\hbar}\,\del_1^2\H(v,\,u)\,-\,4s^2\frac{\rmi}{\hbar}\,X\,\del_1\del_2\H(v,\,u)
\,-\,8s^4\frac{\rmi}{\hbar}X^2\,\del_2^2\H(v,\,u).
\end{eqnarray}

\noindent So this is -- for every choice of boundary perturbation
(\ref{Gl:PropagatorInitialPerturbations}) -- a solution of
(\ref{Gl:PropagatorDGLForX}). The boundary perturbations have to
be chosen in a way to satisfy
(\ref{Gl:PropagatorInitialConditionForX}), which can easily be
done by choosing $\d u(0)=0$,
i.e.~by only perturbing $v(\t)$.\\

 With this knowledge, we are now able to rewrite the
factor (\ref{Gl:PropagatorLimitXFactor}) with the help of
(\ref{Gl:PropagatorEvolutionOfPerturbations}):
\begin{eqnarray}
2s^2\,\frac{\d u}{\d
v}\,\frac{\rmi}{\hbar}\,\del_2^2\H(v,\,u)\;=\;\frac{\d\dot{ v}}{\d
v}\,-\,2s^2\frac{\rmi}{\hbar}\del_1\del_2\H(v,\,u),
\end{eqnarray}

\noindent and therefore
\begin{eqnarray}\nonumber
\fl\exp\left(-4s^4\frac{\rmi}{\hbar}\int_0^{\t}dt\,X\,\del_2^2\H\right)
\;=\;\exp\left(-\frac{\rmi}{2\hbar}\int_0^{\t}dt\,\frac{d}{dt}\,\ln
\d\bar
w\;+\;s^2\frac{i}{\hbar}\int_0^{\t}dt\,\del_1\del_2\H\right)\\[5pt]\label{Gl:PropagatorRewritingOfExponentialWithX}
\;=\;\sqrt\frac{\d v(0)}{\d
v(\t)}\;\exp\left(s^2\frac{\rmi}{\hbar}\int_0^{\t}dt\,\del_1\del_2\H\right).
\end{eqnarray}

\noindent Combining the results
(\ref{Gl:PropagatorLimitSFactorLimit}),
(\ref{Gl:PropagatorLimitXFactor}) and
(\ref{Gl:PropagatorRewritingOfExponentialWithX}), the limit
$N\to\infty$ of
(\ref{Gl:PropagatorBasicFormulaAfterLastIntegration}) becomes:
\begin{eqnarray}\nonumber
\fl{}_{\d}\langle z_{F}|\rme^{-\frac{\rmi}{\hbar}\hat
H\t}|z_I\rangle_{\d}\;\approx\;
\sqrt\frac{\pi}{s^2}\;\sum_{n\in\Z}\;\rme^{2\pi \rmi
n\d}\;\left\{\sqrt\frac{\d v(0)}{\d
v(\t)}\right\}_n\;\exp\left\{s^2\frac{\rmi}{\hbar}\int_0^{\t}dt\,\del_1\del_2\H\right\}_n\\[5pt]\nonumber
\times \; \exp\Bigg\{\int_0^{\t}dt\,\left(\frac{\dot{v}u-\dot u
v}{4s^2}-\frac{\rmi}{\hbar}\H(v,\,u)\right)\;-\;\left(\frac{v(0)-z_I}{2s}\right)^2\\[5pt]\label{Gl:PropagatorNearlyFinalFormula}
\quad-\;\left(\frac{\bar z_F-2\pi n-u(\t)}{2s}\right)^2\;+\;\frac{v(0)\big(v(0)-z_I\big)}{4s^2}\\[5pt]\nonumber
\quad+\;\frac{u(\t)\big(u(\t)-\bar z_F+2\pi
n\big)}{4s^2}\Bigg\}_n.
\end{eqnarray}

\noindent Here the subscript $n$ shall remind us of the fact that
the propagator is a sum over all $n$, where for each $n$ the
complex classical trajectory $u(t),\,v(t)$ given by
(\ref{Gl:PropagatorHamiltonEquationsInUandV}) is different: $u$
starts at $z_I$ and $v$ ends at $\bar z_F-2\pi n$. The function
$X(t)$ for each of these paths has to be computed separately, and
at the end all propagators for these paths have to be summed up,
each one with a phase $\rme^{2\pi \rmi n\d}$.\\

 As a consistency check we consider the case $\t\to 0$.
Then the solution of the Hamiltonian equations
(\ref{Gl:PropagatorHamiltonEquationsInUandV}) become trivial for
every $n\in\Z$, in particular $u(t)=z_I$ and $v(t)=\bar z_F-2\pi
n$. With this, (\ref{Gl:PropagatorNearlyFinalFormula}) becomes:
\begin{eqnarray}
\fl\sqrt\frac{\pi}{s^2}\sum_{n\in\Z}\rme^{2\pi \rmi n \d}\sqrt{1}\;\exp(0)\exp\left[0\,-\,\left(\frac{2\pi n+\bar
z_F-z_I}{2s}\right)^2\,-\,\left(\frac{\bar z_F-2\pi
n-z_I}{2s}\right)^2\right.\\[5pt]\nonumber
\quad\,+\,\left.\frac{(\bar z_F-2\pi n)\big(\bar z_F-2\pi
n-z_I\big)}{4s^2}\,+\;\frac{z_I\big(z_I-\bar z_F+2\pi
n\big)}{4s^2}\right]\\[5pt]\nonumber
\;=\;\sqrt\frac{\pi}{s^2}\sum_{n\in\Z}\exp\left[{2\pi \rmi n \d}\;-\;\left(\frac{\bar z_F-2\pi
n-z_I}{2s}\right)^2\right]\;=\;{}_{\d}\langle z_F|z_I\rangle_{\d},
\end{eqnarray}

\noindent which is the overlap of two coherent states
(\ref{Gl:CoherentOverlap}).

\subsection{The complex action}

\noindent Like in \cite{KECK}, the prefactor with the square-root
in (\ref{Gl:PropagatorNearlyFinalFormula}) can be rewritten in
terms of a complex action. We will do the same here and write, in
accordance to \cite{KECK}, the boundary points of the classical
trajectories as:
\begin{eqnarray}\nonumber
&u'\;:=\;u(0)\;=\;z_I,\qquad
&u''\;:=\;u(\t)\\[5pt]\nonumber
&v'\;:=\;v(0),\qquad &v''\;:=\;v(\t)\;=\;\bar z_F-2\pi n.
\end{eqnarray}

\noindent Then we define the complex action to be
\begin{eqnarray}\label{Gl:PropagatorComplexAction}
\fl
S(u',\,v'',\,\t)\;:=\;\int_0^{\t}dt\;\left[\frac{\rmi\hbar}{4s^2}(\dot
u v - \dot v u)\,-\,\H(u,\,v)\right]\;-\;
\frac{\rmi\hbar}{4s^2}(u'v'+u''v''),
\end{eqnarray}

\noindent where the independent variables in $S$ are $u',\,v''$
and $\t$. The complex trajectories $u(t)$ and $v(t)$ result from
the variational principle with $S$ as complex action and are hence
functions of $u',\,v''$ and $\t$. Perturbing these variables, the
variation of $S$ is given by
\begin{eqnarray}\label{Gl:PropagatorVariationOfS}
\fl\d
S(u',\,v'',\,\t)\;=\;\int_0^{\t}dt\left[\left(\frac{\rmi\hbar}{2s^2}\dot
u-\del_1\H\right)\d v
\;-\;\left(\frac{i\hbar}{2s^2}\dot v-\del_2\H\right)\d u\right]\\[5pt]\nonumber
-\;\frac{\rmi\hbar}{2s^2}\big(v'\d u' + u'' \d
v''\big)\;-\;\H(u'',\,v'',\,\t)\d\t.
\end{eqnarray}

\noindent The integral vanishes, since the classical trajectories
are defined by the solutions of Hamilton's equations
(\ref{Gl:PropagatorHamiltonEquationsInUandV}). Hence we get
\begin{eqnarray}\label{Gl:PropagatorDerivativesOfS}
\fl\frac{\del S}{\del
u'}\;=\;-\frac{i\hbar}{2s^2}v',\qquad\frac{\del S}{\del
v''}\;=\;-\frac{i\hbar}{2s^2}u'',\qquad \frac{\del S}{\del
\t}\;=\;-\H(u'',\,v'',\,\t).
\end{eqnarray}

\noindent The first equation of
(\ref{Gl:PropagatorDerivativesOfS}) shows that
\begin{eqnarray}
\frac{\del^2S}{\del u'\del
v''}\;=\;-\frac{\rmi\hbar}{2s^2}\frac{\d v'}{\d v''}
\end{eqnarray}

\noindent and (\ref{Gl:PropagatorNearlyFinalFormula}) can be
rewritten in terms of the complex action $S$, which yields the
final result for the coherent state propagator
(\ref{Gl:CoherentPropagator}) in the semiclassical approximation:
\begin{eqnarray}\nonumber
\fl K_{S^1}\big(z_F,\,\t,\,z_I,\,0\big)\;\approx\;
\sqrt{2\pi}\;\sum_{n\in\Z}\sum_{\n}\;e^{2\pi \rmi
n\d}\;\sqrt{\frac{\hbar}{\rmi}\frac{\del^2S_{\n}}{\del
u'\del v''}\big(z_I,\,\bar z_F-2\pi n,\,\t\big)}\\[5pt]\label{Gl:PropagatorFinalFormulaCoherent}
\times \;\exp\left\{s^2\frac{\rmi}{\hbar}\int_0^{\t}dt\,\del_1\del_2\H\right\}_{n,\n}\\[5pt]\nonumber
\times \; \exp\left\{\frac{\rmi}{\hbar}S_{\n}\big(z_I,\,\bar
z_F-2\pi n,\,\t\big) \;-\;\frac{z_I^2+(\bar z_F-2\pi
n)^2}{4s^2}\right\}.
\end{eqnarray}

\noindent Here the sum over $\n$ shall indicate that for each $n$
there may be different complex paths $u$ and $v$ satisfying
(\ref{Gl:PropagatorHamiltonEquationsInUandV}) and
(\ref{Gl:PropagatorHamiltonEquationsBoundaryInUandV}). Each of
them has to be computed separately and taken into account in
formula (\ref{Gl:PropagatorFinalFormulaCoherent}). \\

 By comparing the result
(\ref{Gl:PropagatorFinalFormulaCoherent}) for the propagator
$K_{S^1}$ on the circle with the coherent state propagator
$K_{\R}$ for quantum mechanics on the real line \cite{KECK}, one
readily sees
\begin{eqnarray}\label{Gl:PropagatorCoherentSuperposition}
K_{S^1}\big(z_F,\,\t,\,z_I,\,0\big)\;\sim\;\sum_{n\in\Z}\rme^{2\pi
n\d \rmi}\,K_{\R}\big(z_F-2\pi n,\,\t,\,z_I,\,0\big).
\end{eqnarray}

\noindent This demonstrates that quantum mechanics on the circle
is nothing but a periodic quantum mechanic on the real line, but
with a phase shift for each period. The proportionality factor has
to be chosen to ensure that both propagators are normalized
correctly.

\subsection{Angle representation}

\noindent From formula (\ref{Gl:PropagatorFinalFormulaCoherent})
on can now obtain, for instance, the angle representation
$\langle\phi_{F}|\rme^{-\frac{\rmi}{\hbar}\hat H\t}|\phi_I\rangle$
by convolution with the coherent states as functions of $\phi$. We
just state the final result:

\begin{eqnarray}\label{Gl:PropagatorPROPAGATOR}
\fl\langle\phi_{F}|\rme^{-\frac{\rmi}{\hbar}\hat
H\t}|\phi_I\rangle\;\approx\;
\sqrt\frac{2\pi}{s^2}\;\sum_{n\in\Z}\sum_{\n}\;\rme^{2\pi \rmi
n\d}\;\frac{1}{\sqrt{(m_{qp})_{n,\n}}}\\[5pt]\nonumber
\times
\exp\left\{s^2\frac{\rmi}{\hbar}\int_0^{\t}dt\,\Delta\H(\phi,p)\right\}_{n,\n}\;
\exp\left\{{\frac{\rmi}{\hbar}\int_0^{\t}dt\,\frac{\dot\phi
p}{s^2}-\H(\phi, \,p)}\right\}_{n,\n},
\end{eqnarray}

\noindent with

\begin{eqnarray}
\H(\phi,\,p)\;=\;\frac{{}_{\d}\Big\langle\phi+{\rmi}p/{\hbar}\,\Big|\,\hat
H\,\Big|\phi+{\rmi}p/{\hbar}\Big\rangle_{\d}}{{}_{\d}\Big\langle\phi+{\rmi}p/{\hbar}\,\Big|\phi+{\rmi}p/{\hbar}\Big\rangle_{\d}}.
\end{eqnarray}

\noindent The integrals in (\ref{Gl:PropagatorPROPAGATOR}) have to
be taken over the solutions of the classical trajectories
satisfying
\begin{eqnarray}\nonumber
\dot\phi\;=\;{2s^2}\frac{\del\H}{\del
p}\left(\phi,\,{p}\right),\qquad \dot
p\;=\;-{2s^2}\frac{\del\H}{\del\phi}\left(\phi,\,p\right)\\[5pt]\label{Gl:PropagatorRealHamiltonEquations}
\phi(0)\;=\;\phi_I,\qquad\phi(\t)\;=\;\phi_F-2\pi n,
\end{eqnarray}

\noindent and $m_{qp}$ is an entry of the tangent matrix
\cite{KECK} that can be computed from the complex action
(\ref{Gl:PropagatorComplexAction}). The additional index $\n$
 indicates that, as already mentioned, even for paths with
fixed $n$, there may be more than one classical trajectory
satisfying (\ref{Gl:PropagatorRealHamiltonEquations}). If there
is, all have to be taken into account. In the angle
representation, the sum over $n$ has a nice interpretation: On a
circle, a particle can go from $\phi_I$ to $\phi_F$ in infinitely
many ways: the different paths can all differ by their relative
winding number. The parameter $\d$ determines the phaseshift the
particle acquires by "going round the circle". All these paths
contribute to the propagator, each one with an additional factor
of $\rme^{2\pi \rmi n \d}$. This  demonstrates the fact that the
motion of a quantum mechanical particle depends on the global
topology of the space it is moving in. The propagator
(\ref{Gl:PropagatorPROPAGATOR}) takes all these paths into account
correctly.

 The mathematical realization of this becomes more
transparent, when comparing (\ref{Gl:PropagatorPROPAGATOR}) to the
result for the propagator on the real line \cite{KECK}:
\begin{eqnarray}\label{Gl:PropagatorAngleSuperposition}
K_{S^1}\big(\phi_F,\,\t,\,\phi_I,\,0\big)\;\sim\;\sum_{n\in\Z}\rme^{2\pi \rmi n \d}\,K_{\R}\big(\phi_F-2\pi n,\,\t,\,\phi_I,\,0\big).
\end{eqnarray}

\noindent Like in (\ref{Gl:PropagatorCoherentSuperposition}), the
proportionality factor has to be chosen to normalize both
propagators correctly.

\section{Summary and conclusion}

\noindent A brief overview of the complexifier coherent states for
quantum mechanics on the circle has been given, summarizing the
results from various authors. Furthermore it was shown that these
states are useful for semiclassical analysis, by considering the
Bargmann-Segal representation and the semiclassical propagator.

The Bargmann-Segal representation, from which the Husimi
distribution can be defined, shows significant differences
compared to the one for quantum mechanics on the real line.
Periodicity of the system restricts the possible forms of the
phase-space wavefunctions $\psi_{\d}(z)$. In particular, apart
from the zeros of $\psi_{\d}(z)$, the state is completely
determined by the choice of normalization and an additional integer. This
is in contrast to the situation on the real line, where, apart
from the zeros of the phase space wavefunction $\psi(z)$, one has
to specify normalization and two arbitrary complex coefficients to
reconstruct the state. Furthermore, on the real line, the coherent
states are the only states whose Husimi distribution has no zeros
at all. On the circle, the eigenvectors $|n\rangle_{\d}$ of the
momentum operator $\hat p$ are the only vectors whose Husimi
distribution is without zeros. This demonstrates the qualitatively
different behavior of the two kind of coherent states, which
results from the different topologies of phase space.

The semiclassical propagator for the complexifier coherent states
was derived. The result is the infinite sum over coherent state
propagators on the real line, each one evaluated for a path with a
different winding number. Also, each of the single propagators
acquires a phase which is determined by the parameter $\d$ of the
representation. This shows the influence of the global phase space
topology on the motion of a quantum mechanical particle: In
contrast to the real line, a wavefunction moving on a circle can
interfere with itself because of the non-simply connected
configuration space. The result also illustrates the physical
meaning of the representation parameter $\d$: it determines the
phase-shift the wavefunction acquires when moving around the
circle. The parameter $\d$ can be chosen according to the problem
at hand and has to be taken into account when computing the
coherent state propagator.

These results show that the complexifier coherent states for
$G=U(1)$ are in principle as useful for a semiclassical analysis of
periodic systems as the harmonic oscillator coherent states are
for systems on the real line.

\ack

BB would like to thank Thomas Thiemann for vivid discussions about
the semiclassical propagator. The authors would like to thank Hans
Kastrup for pointing out the significance of reference
\cite{KASTRUP}.

\appendix

\section{Hadamard-decomposition of the sine}

\noindent In this section, we will prove formula
(\ref{Gl:HadamardOfSin}).  We start by noticing that the sine
function is of first order:
\begin{eqnarray}
\left|\sin\frac{z-a}{2}\right|\;\leq\;\exp\left({\frac{1}{2}|a|}\,+\,{\frac{1}{2}|z|}\right).
\end{eqnarray}

\noindent So, the function $z\mapsto \sin\frac{z-a}{2}$ can,
according to the Hadamard-decomposition, be written \cite{RC95} in
the following way:
\begin{eqnarray}\nonumber
\fl\sin\frac{z-a}{2}=\rme^{C_0+C_1z}\;\prod_{n\in\Z}\left(1-\frac{z}{a+2\pi
n}\right)\exp\left({\frac{z}{a+2\pi n}}\right)\\[5pt]\nonumber
\fl=\rme^{C_0+C_1z}\left(1\!-\!\frac{z}{a}\right)\,e^{\frac{z}{a}}\prod_{n=1}^{\infty}\left(1\!-\!\frac{z}{a\!+\!2\pi
n}\right)\left(1\!-\!\frac{z}{a\!-\!2\pi
n}\right)\exp\left(\frac{z}{a\!+\!2\pi
n}\!+\!\frac{z}{a\!-\!2\pi n}\right)\\[5pt]
\fl=\rme^{C_0+\widetilde{
C_1}z}\;\left(1-\frac{z}{a}\right)\prod_{n=1}^{\infty}\left(1-\frac{z}{a+2\pi
n}\right)\left(1-\frac{z}{a-2\pi n}\right),
\end{eqnarray}

\noindent with
\begin{eqnarray}
\widetilde{
C_1};:=\;C_1\;+\;\frac{1}{a}\;+\;\sum_{n=1}^{\infty}\frac{2a}{a^2-4\pi^2n^2}.
\end{eqnarray}

\noindent By setting $z=0$, one gets the condition
$\rme^{C_0}=-\sin\frac{a}{2}$, and from the anti-periodicity of
the sine,
\begin{eqnarray}\nonumber
\fl-1\;=\;\frac{\sin\frac{z+2\pi-a}{2}}{\sin\frac{z-a}{2}}\;&=\;\rme^{2\pi\widetilde{
C_1}}\;\prod_{n\in\Z}\frac{1-\frac{z+2\pi}{a+2\pi
n}}{1-\frac{z}{a+2\pi n}}\\[5pt]
&=\;\rme^{2\pi\widetilde{
C_1}}\lim_{N\to\infty}\prod_{n=-N}^N\frac{a+2\pi(n-1)-z}{a+2\pi n-z}\\[5pt]\nonumber
&=\;\rme^{2\pi\widetilde{
C_1}}\lim_{N\to\infty}\frac{a-2\pi(N+1)-z}{a+2\pi N-z}\\[5pt]\nonumber
&=\;-\rme^{2\pi\widetilde{
C_1}}.
\end{eqnarray}

\noindent Thus, $\rme^{2\pi\widetilde{
C_1}}=1$ and $\widetilde{
C_1}=\rmi k$,
with $k$ an integer. This means that

\begin{eqnarray}\label{Gl:Appendix1}
\fl\sin\frac{z-a}{2}\;=\;-\sin\frac{a}{2}\,\rme^{\rmi k
z}\,\left(1-\frac{z}{a}\right)\,\prod_{n=1}^{\infty}\left(1-\frac{z}{a+2\pi
n}\right)\left(1-\frac{z}{a-2\pi n}\right)
\end{eqnarray}

\noindent We will now show that $k$ is actually zero. To achieve
this, we set $z=2a$, and obtain:
\begin{eqnarray}
\sin\frac{a}{2}\;=\;\sin\frac{a}{2}\;\rme^{2\rmi k a
}\;\prod_{n=1}^{\infty}\left(1-\frac{2a}{a+2\pi
n}\right)\left(1-\frac{2a}{a-2\pi n}\right)
\end{eqnarray}

\noindent and therefore

\begin{eqnarray}\nonumber
\rme^{-2\rmi k a }\;&=\;\prod_{n=1}^{\infty}\left(1-\frac{2a}{a+2\pi n}\right)\left(1-\frac{2a}{a-2\pi n}\right)\\[5pt]
&=\;\prod_{n=1}^{\infty}\frac{2\pi n-a}{2\pi n+a}\,\frac{-2\pi
n-a}{a-2\pi n}\;=\;1,
\end{eqnarray}

\noindent which immediately shows that, if $k\neq 0$, $a=\pi m/k$ for some $m\in Z$.  But then the function
$z\mapsto\sin\frac{z-a}{2}$ is real for $z\in\R$, which, as can be
seen by (\ref{Gl:Appendix1}), is only possible for $k=0$. So we
have

\begin{eqnarray}\label{Gl:Appendix2}
\fl\sin\frac{z-a}{2}\;=\;-\sin\frac{a}{2}\,\left(1-\frac{z}{a}\right)\,\prod_{n=1}^{\infty}\left(1-\frac{z}{a+2\pi
n}\right)\left(1-\frac{z}{a-2\pi n}\right)
\end{eqnarray}

\noindent which is formula (\ref{Gl:HadamardOfSin}).\\

\section{Proof of a technical lemma}
\label{app-proof}
\noindent In this section, the proof for
Lemma
\ref{Lem:EinzigesLemmaImArtikel} will be delivered.\\

\noindent {\it Let $\{a_k\}$ be a sequence in $([0,2\pi)\times
\rmi\R)\backslash\{0\}$ of order 1. Let $\n_k\,:=\,{\rm sgn}\,{\rm
Im}\,a_k$ be the sign of the imaginary part of $a_k$ (with {\rm
sgn}\,0\;:=\;1). Then}
\begin{eqnarray}
\prod_k\Bigg[-\exp\Big({\pi\cot\frac{a_k}{2}}\Big)\Bigg]\;=\;\exp\left[\sum_k\pi\left(
\cot\frac{a_k}{2}\,+\,\n_k\rmi\right)\right]
\end{eqnarray}\\

\noindent\textbf{Proof:} We would like to employ the formula
\begin{eqnarray}\label{Gl:ExponentialWithLogarithm}
\prod_kf_k\;=\;\exp\left[\sum_k\ln\;f_k\right].
\end{eqnarray}

\noindent To use this formula for complex numbers $f_k$, one has
to select  a branch cut of the logarithm.
The particular choice of
the cut results in different imaginary parts of (some of the)
$\ln\,f_k$, differing by an integer multiple of $2\pi \rmi$. The
exponential of these different $\ln\,f_k$ hence is the same for
all choices of branch cut, which means that
(\ref{Gl:ExponentialWithLogarithm}) is valid for every choice of
branch cut of the logarithm.
In the following, we choose the cut on
the negative real axis, i.e.
\begin{eqnarray}\label{Gl:DefinitionOfBranchCut}
\ln\,\exp\left(\rmi x\right)\;=\;\left\{\begin{array}{ll} &\vdots\\
\rmi x+2\pi\rmi&x\in(-3\pi,-\pi]\\\rmi x & x \in(-\pi,\pi]
\\ \rmi x-2\pi\rmi & x\in(\pi,3\pi]\\&\vdots\end{array}\right.\,.
\end{eqnarray}

\noindent So, we have to compute $\ln(-\exp(\pi\cot\,a_k/2))$. The
result crucially depends on the imaginary part of $\a_k$, so we
need to keep track of it.

Since the sequence $\{a_k\}_k$ is of order 1, it has no
accumulation points. So, since $a_k\in[0,2\pi)\times \rmi\R$,
there is only a finite number of $a_k$ with fixed real part. Thus we
can order the sequence $\{a_k\}_k$ by ascending imaginary part,
i.e.
\begin{eqnarray}
{\rm Im}\,a_k\,\leq\,{\rm Im}\,a_l\;\Leftrightarrow\;k<l.
\end{eqnarray}

\noindent
The number of the  $a_k$ can be finite or infinite,
which leaves three possibilities: The index
$k$ ranges from $-\infty$ to $0$, from $0$ to $\infty$, or from
$-\infty$ to $\infty$, depending on
the distribution of the $a_k$ over the strip.
We will concentrate on the last of these four
cases, which is the most general one. The argument for the other
three cases runs along similar lines.\\

So, the $a_k$ go to $\rmi\infty$ for $k\to\infty$ and to
$-\rmi\infty$ for $k\to\-\infty$. Since ${\rm
Re}\;a_k\in[0,2\pi)$, we have
\begin{eqnarray}
\lim_{k\to \pm\infty}\,{\rm Im}\cot\frac{a_k}{2}\;=\;\mp 1\,,
\end{eqnarray}
which follows from the properties of the
cotangent.
So there are $N,M\in\Z$ such that
\begin{eqnarray}
{\rm
Im}\,\cot\frac{a_k}{2}\,\in\,{\textstyle\left( \frac{1}{2},\,\frac{3}{2}\right)},&\quad{\rm
Im}\;a_k\;<\;0\qquad\mbox
{\rm for all }k\,<\,N,\\[5pt]
{\rm
Im}\,\cot\frac{a_k}{2}\,\in\,{\textstyle\left(-\frac{3}{2},\,-\frac{1}{2}\right)},&\quad{\rm
Im}\;a_k\;>\;0\qquad\mbox{\rm for all }k\,>\,M.
\end{eqnarray}

\noindent Hence, with the definition of the
logarithm
(\ref{Gl:DefinitionOfBranchCut}), we have
\begin{eqnarray}
\ln\left[-\exp\left(\pi\,\cot\frac{a_k}{2}\right)\right]\;=\;\left\{\begin{array}{ll}
\pi \cot\frac{a_k}{2}\,-\,\rmi\pi, & \quad k<N\\[5pt]
\pi \cot\frac{a_k}{2}\,+\,\rmi\pi, & \quad k>M
\end{array}\right.\,.
\end{eqnarray}

\noindent This means that, with $\n_k:={\rm sgn}\,{\rm Im}\,a_k$,:
\begin{eqnarray}
\ln\left[-\exp\left(\pi\,\cot\frac{a_k}{2}\right)\right]\;=\;\pi\left(
\cot\frac{a_k}{2}\,+\,\rmi\n_k\right),
\end{eqnarray}

\noindent for all $k$, up to finitely many exceptions. So, there
is an integer $K\in\Z$, such that
\begin{eqnarray}
\sum_{k\in\Z}\ln\left[-\exp\left(\pi\,\cot\frac{a_k}{2}\right)\right]\;=\;\sum_{k\in\Z}\pi\left(
\cot\frac{a_k}{2}\,+\,\rmi\n_k\right)\;+\;2\pi\rmi K.
\end{eqnarray}

\noindent From this it follows immediately that
\begin{eqnarray}
\fl \prod_{k\in\Z}\Bigg[-\exp\Big({\pi\cot\frac{a_k}{2}}\Big)\Bigg]
=\exp\sum_{k\in\Z}\ln\left[-\exp\Big({\pi\cot\frac{a_k}{2}}\Big)\right]\nonumber \\[5pt]
\fl \qquad = \exp\left[\sum_k\pi\left(
\cot\frac{a_k}{2}\,+\,\n_k\rmi\right)\;+\;2\pi\rmi K\right]
=\exp\sum_k\pi\left( \cot\frac{a_k}{2}\,+\,\n_k\rmi\right),
\end{eqnarray}
as claimed in the lemma.

Note that the Lemma stays true, of course, if one replaces
$\n_k={\rm sgn}\,{\rm Im}\,a_k$ by any other sequence
$\tilde\n_k\in\{\pm1,\,\pm3,\,\pm5,\ldots\}$, that differs with
$\n_k$ at at most finitely many $k$'s. Only the sum is shifted by
an integer multiple of $2\pi\rmi$, which does not show up in the
exponential.

But in the case of infinitely many $a_k$'s, the choice of the
$\n_k={\rm sgn}\,{\rm Im}\,a_k$ (up to finitely many exceptions)
is important for the sum to converge, that is, one is not allowed
to change the $\n_k$ at more that finitely many $k$'s, in order
for the sum $\sum_k\pi\left(
\cot\frac{a_k}{2}\,+\,\n_k\rmi\right)$ to exist at all. The proof
that the sum then actually converges rests on the fact that the
sequence $\{a_k\}$ is of order 1.

The lemma is only true if we define the sign of 0 to be either $1$
or $-1$. For real $a_k$, for which the imaginary part of
$\cot\,a_k/2$ vanishes, one has
\begin{eqnarray}
\ln\left[-\exp\left(\pi\cot\frac{a_k}{2}\right)\right]\;=
\;\pi\left(\cot\frac{a_k}{2}\;\pm\;\rmi\right),
\end{eqnarray}

\noindent the sign depending on the particular choice of the
branch cut. So, for real $a_k$, one has to have either $\n_k=+1$
of $\n_k=-1$, although ${\rm sgn}\,{\rm Im}\,a_k=0$. Thus, we
adjust the definition of the sign of $0$ to be $+1$, and with this
definition the lemma is thoroughly true.  Since there are only
finitely many real $a_k$, we could have chosen the sign of $0$ to
be $-1$ as well, for the reasons stated above.\\[10pt]

\section*{References}

\end{document}